\documentclass[aps,pra,nofootinbib,twocolumn,floatfix,showpacs,superscriptaddress]{revtex4-1}
\usepackage[utf8]{inputenc}
\usepackage{amsmath,graphicx,epsfig,amsthm,color,bm}
\usepackage{pifont,bbding,amssymb,soul}
\usepackage{indentfirst}
\usepackage{times}
\usepackage{appendix, comment}

\usepackage{CJKutf8}
\usepackage{ucs}
\usepackage[encapsulated]{CJK}

\usepackage{tikz,pgfplots}

\newcommand{\braket}[1]{\langle #1 \rangle}
\newcommand{\proj}[1]{\ket{#1}\!\bra{#1}}
\newcommand{\tr}{{\rm tr}}

\newcommand{\bra}[1]{\langle #1|}
\newcommand{\ket}[1]{|#1\rangle}

\newtheorem{theorem}{Theorem}

\newcommand{\id}{\mathbb{I}}
\usepackage[colorlinks=true,citecolor=blue,urlcolor=blue]{hyperref}

\usepackage[capitalize]{cleveref}

\begin{document}

\title{Efficient estimation of multipartite quantum coherence}

\author{Qi-Ming Ding}
\email{These authors contributed equally to this work}
\affiliation{School of Physics, Shandong University, Jinan 250100, China}

\author{Xiao-Xu Fang}
\email{These authors contributed equally to this work}
\affiliation{School of Physics, Shandong University, Jinan 250100, China}

\author{Xiao Yuan}
\affiliation{Stanford Institute for Theoretical Physics, Stanford University, Stanford California 94305, USA}

\author{Ting Zhang}
\affiliation{School of Physics, Shandong University, Jinan 250100, China}

\author{He Lu}
\email{luhe@sdu.edu.cn}
\affiliation{School of Physics, Shandong University, Jinan 250100, China}

\begin{abstract}
Quantification of coherence lies at the heart of quantum information processing and fundamental physics. Exact evaluation of coherence measures generally needs a full reconstruction of the density matrix, which becomes intractable for large-scale multipartite systems. Here, we propose a systematic theoretical approach to efficiently estimating lower and upper bounds of coherence in multipartite states. Under the stabilizer formalism, the lower bound is determined by the spectrum estimation method with a small number of measurements and the upper bound is determined by a single measurement. We verify our theory with a four-qubit optical quantum system.  
We experimentally implement various multi-qubit entangled states, including the Greenberger-Horne-Zeilinger state, the cluster state, and the W state, and show how their coherence are efficiently inferred from measuring few observables.  
\end{abstract}

\maketitle

\section{Introduction}
Quantum coherence, being one of the defining features of quantum mechanics, underlies the fundamental phenomena of quantum interference and plays a significant role in physics and quantum information processing (QIP), such as quantum cryptography~\cite{Grosshans03,coles2016numerical,ma2019operational}, quantum metrology~\cite{Giovannetti2011,zhang2019demonstrating}, nanoscale thermodynamics~\cite{aaberg2014catalytic,Lostaglio2015, Narasimhachar2015}, and energy transport in biological system~\cite{Romero2014}. Based on the general framework of quantum resource theories~\cite{Vedral1997, Vedral1998, RevModPhys.91.025001}, a systematic framework of coherence have been introduced~\cite{aberg2006, Baumgratz2014},  based on which various coherence measures have been defined~\cite{Baumgratz2014,streltsov_measuring_2015,napoli_robustness_2016,bu_maximum_2017,chen_notes_2018,du_coherence_2019,du_coherence_2015,liu_new_2017,qi_measuring_2017,xi_coherence_2019,xi_epsilon-smooth_2019,zhao_one-shot_2018,zhou_polynomial_2017}. 
Meanwhile, the framework of coherence has been extended from a single party to the multipartite scenario with several applications, such as quantum state merging~\cite{streltsov_entanglement_2016}, coherence localization~\cite{chitambar_assisted_2016} and incoherence teleportation~\cite{streltsov_towards_2017}. Studies of the inter-conversion between coherence and other multipartite nonclassical correlations, such as entanglement~\cite{streltsov_measuring_2015,Chitambar_coherence-entanglement2016,Girolami2017}, discord~\cite{ma_converting_2016} and nonlocality~\cite{killoran_converting_2016}, also highlight the fundamental role of quantum coherence. With the rapid development of quantum hardware in realizing large-scale multipartite systems, the ability of efficiently quantifying the coherence would thus offer an operationally meaningful benchmarking tool and benefit our understanding of QIP tasks. 

Several experiments have been reported regarding the efficient detection of robustness of coherence~\cite{wang_directly_2017, zheng_experimental_2018}. However, the experimental detection of general coherence measure, such as the relative entropy of coherence~\cite{Baumgratz2014}, is still missing. Theoretical proposals to estimate general multipartite coherence without costly state tomography have also been proposed~\cite{smith_quantifying_2017,zhang_estimating_2018,yu_detecting_2019}.
While the initial proposals either need copies of the prepared multipartite state~\cite{smith_quantifying_2017} or complicated post-processing ~\cite{zhang_estimating_2018}, the spectrum estimation method was recently proposed~\cite{yu_detecting_2019}, which only requires local measurements and easy-to-compute post-processing. 
Nevertheless, the performance of the spectrum estimation method highly depends on the choice of the measurements, and how it works for a general multipartite state still needs further study. 
Moreover, existing works generally focus on the lower bound of coherence. For a given quantum state, the maximal entanglement or discord it can generates are upper bounded by its coherence~\cite{streltsov_measuring_2015, ma_converting_2016}, which makes detecting the upper bound of coherence important as it can indicates whether the given quantum state can generate sufficient resource for a certain QIP task. 

In this work, we theoretically address these issues by proposing two methods that can respectively detect the lower and upper bound of coherence for all multi-qubit stabilizer states. The lower bound detection is based on the spectrum estimation method~\cite{yu_detecting_2019} and the stabilizer theory~\cite{Gottesman1996,GottesmanPhD1997}, which only requires few local observable measurements for stabilizer states. The upper bound detection is based on the monogamy of coherence with a single local measurement. Experimentally, we  prepare five stabilizer states of up to four qubits and demonstrate how few number of measurements could enable us to infer multipartite coherence.

\section{Theory}
\subsection{Lower bound estimation}
Under the computational basis $\{\ket{i}:i\in\{0,1\}^{\otimes n}\}$ of an $n$-qubit state, we consider the relative entropy of coherence~\cite{Baumgratz2014} 
\begin{equation}\label{Eq:RE}
    C_{\text{RE}}(\rho)=S_{\text{VN}}(\rho_{d})-S_{\text{VN}}(\rho),
\end{equation}
with  $S_{\text{VN}}=-\tr[\rho\log_2\rho]$ being the von Neumann entropy and $\rho_d = \sum_{i}\braket{i|\rho|i}\ket{i}\bra{i}$ being the diagonal part of $\rho$. {The relative entropy of coherence characterizes the asymptotic distillable coherence under different types of incoherent operations~\cite{winter_operational_2016, zhao2019one}, quantifies the genuine randomness that can be extracted from measuring the quantum state in the computational basis~\cite{yuan_intrinsic_2015,hayashi2018secure,yuan2019quantum},  captures the deviation from thermodynamic equilibrium~\cite{rodriguez2013thermodynamics}, etc. We thus focus on the estimation, in particular, the lower and upper bounds, of the relative entropy of coherence for general multipartite states.}

The lower bound $l^c(\rho)$ of the coherence $C_{\text{RE}}(\rho)$ can be obtained by spectrum estimation and the majorization theory~\cite{cicalese_supermodularity_2002} as
\begin{equation}\label{eq:MultpartiteEstimation}
   C_{\text{RE}}(\rho)\geq l^{c}(\rho)=S_{\text{VN}}(\bm{d})-S_{\text{VN}}(\bm{d} \vee (\wedge_{\bm{p} \in X}\bm{p})),
\end{equation}
where $\bm{d}=(d_1,...,d_{2^n})$ are the diagonal elements of $\rho$, $\bm{p}=(p_1,...,p_{2^n})$ is the estimated probability distribution of the measurement on a certain entangled basis $\{\ket{\psi_k}\}_{k=1}^{2^n}$, $\vee$ is majorization joint, and $\wedge_{\bm{p} \in X}\bm{p}$ is the majorization meet of all probability distributions in $X$~\cite{yu_detecting_2019}. {Here the majorization join and meet are defined based on majorization. Specifically, given two probability distributions $\boldsymbol{a}=\left(a_{1}, a_{2}, \ldots, a_{n}\right)$ and $\boldsymbol{b}=\left(b_{1}, b_{2}, \ldots, b_{n}\right)$ with $a_{1}\geq a_{2} \geq \ldots \ge a_{n}$ and  $b_{1}\geq b_{2} \geq \ldots \ge b_{n}$, $\boldsymbol{a}$ is majorized by $\boldsymbol{b}$ (written as $\boldsymbol{a} \prec \boldsymbol{b}$) if it satisfies $\sum_{i=1}^{k} a_{i} \leq \sum_{i=1}^{k} b_{i}$ for all $k= 1,2, \ldots, n$. A probability distribution $\boldsymbol{c}$ is called the majorization join (meet) of $\boldsymbol{a}$ and $\boldsymbol{b}$ if it satisfies: (i) $c \succ$ $\boldsymbol{a}, \boldsymbol{b}~(\boldsymbol{c} \prec \boldsymbol{a}, \boldsymbol{b})$, and (ii) $c \prec \tilde{\boldsymbol{c}}~(\boldsymbol{c} \succ \tilde{\boldsymbol{c}})$ for any $\tilde{\boldsymbol{c}}$ that satisfies
$\boldsymbol{a}, \boldsymbol{b} \prec \tilde{\boldsymbol{c}}~(\boldsymbol{a}, \boldsymbol{b} \succ \tilde{\boldsymbol{c}})$}~\cite{cicalese_supermodularity_2002}.
Here, we consider $\bm{p}$ is selected from the set  $X$, which satisfies $X=\{\bm{p}| A \bm{p} \geq \bm{\alpha}, B\bm{p}= \bm{\beta} \}$. $A$ and $B$ are matrices and $\bm{\alpha}$ and $\bm{\beta}$ are vectors. ``$\geq$'' represents component-wise comparison. 

To calculate $l^c(\rho)$ via Eq.~\ref{eq:MultpartiteEstimation}, it is crucial to set constraints $A\bm{p}\geq\bm{\alpha}$ and $B\bm{p}=\bm{\beta}$ with experimentally collected data, then find the ``largest" distribution $\bm{p}$ majorized by all probability distributions in $X$, i.e., $\wedge_{\boldsymbol{p} \in X} \boldsymbol{p}$. According to the hermiticity of density matrix $\rho$, we can set $A$ as \{$2^{n}$-dimensional identity matrix and $\bm{\alpha}=0$ for $A\bm{p}\geq\bm{\alpha}$, by which $p_k\geq0$ is guaranteed. \{In the following section, we introduce the procedure to construct constraint $B\bm{p}=\bm{\beta}$ via stabilizer formalism.

\subsection{Constructing constraint via stabilizer formalism} 
An observable $S_i$ stabilizes an $n$-qubit state $\ket{\psi}$ if $\ket{\psi}$ is an eigenstate with eigenvalue $+1$ of $S_i$, i.e., $S_i\ket{\psi}=\ket{\psi}$. The set $\mathcal S$  of operators $S_i$ is the stabilizer of $\ket{\psi}$, and $\ket{\psi}$ is a so-called a stabilizer state~\cite{GottesmanPhD1997,nielsen_quantum_2010}. For an $n$-qubit state, there are $n$ stabilizing operators $\{S_1,...,S_n\}$ that can uniquely determine $\ket{\psi}$. Here $S_1,...,S_n$ are the generators of the set $\mathcal{S}$, and we denote $\mathcal{S}=\langle S_1,...,S_n\rangle$. Note that $\ket{\psi}$ is not only stabilized by $\{S_1,...,S_n\}$, but also their products. Thus, there could be in total  $2^n$ stabilizer operators in $\mathcal{S}$. 

Given an $n$-qubit stabilizer state $\ket{\psi_1}$ associated with stabilizer $\mathcal{S}$, there exists an orthonormal basis $\{\ket{\psi_k}\}_{k=1}^{2^n}$ including $\ket{\psi_1}=\ket{\psi}$, where $\ket{\psi_k}$ is uniquely specified by $\mathcal{S}$ but with different eigenvalues, i.e., $S_i\ket{\psi_k}=a_{ik}\ket{\psi_k}$ with eigenvalues $a_{ik}=\pm1$. For example, the Bell state $\ket{\Phi^+}=(\ket{00}+\ket{11})/\sqrt{2}$ can be specified by $\mathcal{S}=\langle X^{(1)}X^{(2)},Z^{(1)}Z^{(2)}\rangle$. Hereafter, $X^{(j)}$, $Y^{(j)}$ and $Z^{(j)}$ denote the Pauli matrices $\sigma_{x}$, $\sigma_{y}$ and $\sigma_{z}$ acting on the $j$-th qubit. In the following, qubit index $j$ may be omitted if there is no confusion. The four Bell states $\ket{\Phi^{\pm}}$ and $\ket{\Psi^{\pm}}$ are specified by $\mathcal{S}=\langle XX,ZZ\rangle$ as well, with eigenvalues of $(+1,+1)$, $(-1,+1)$, $(+1,-1)$ and $(-1,-1)$ respectively. The basis $\{\ket{\psi_k}\}_{k=1}^{2^n}$ associated with the same stabilizer $\mathcal{S}$ is also called graph-diagonal basis~\cite{Dur2003,Otfried2011} as stabilizer state is equivalent to a graph state under local Clifford
operations~\cite{schlingemann_stabilizer_2001}. In the graph-diagonal basis, the stabilizing operator can be written as $S_i=\sum_{k} a_{ik} \ket{\psi_k}\bra{\psi_k}$, and its expected value on a given quantum state $\rho$ is $\braket{S_i}=\tr(S_i\rho)=\sum_k p_k a_{ik}$,
where the parameters $p_k=\bra{\psi_{k}} \rho \ket{\psi_{k}}$ form a probability distribution $\bm{p}=(p_1,...,p_{2^n})$ with $p_k\geq0$ and $\sum_kp_k=1$.

The expected values of stabilizer $\mathcal{S}$ on $\rho$ leads to $2^n$ equations, and can be represented in matrix form  
\begin{equation}\label{eq:constraintEq}
    \underbrace{\begin{bmatrix}
    a_{11} &\hdots &a_{12^n} \\ \vdots & \ddots &\vdots\\ a_{2^n1} & \hdots & a_{2^n2^n}
    \end{bmatrix}}_{B}\cdot
    \underbrace{\begin{bmatrix}
    p_1 \\ \vdots \\ p_{2^n}
    \end{bmatrix}}_{\bm{p}}
    =\underbrace{\begin{bmatrix}
   \braket{S_1} \\ \vdots \\ \braket{S_{2^n}}
    \end{bmatrix}}_{\bm{\beta}},
\end{equation}
from which we can construct the constraint $B\bm{p}=\bm{\beta}$. However, in practice, there does not always exist solutions of {$\wedge_{\boldsymbol{p} \in X} \boldsymbol{p}$} in Eq.~\ref{eq:MultpartiteEstimation} {with constraint Eq.~\ref{eq:constraintEq}} (as reflected by our experimental results). {The experimentally generated state always has a distance to target state due to the inevitable imperfections, which might lead to no solution of $\wedge_{\boldsymbol{p} \in X} \boldsymbol{p}$ with inputs of $\{\braket{S_i}\}$. On the other hand, experimentally obtained $\braket{S_i}$ is always associated with statistical errors.} We address these issues by introducing the experimental standard deviation $\sigma_i$ of $\braket{S_i}$ and relax the constraint Eq.~\ref{eq:constraintEq} to an inequality form of 
\begin{equation}\label{Eq:constraintInaeq}
    \underbrace{\begin{bmatrix}
    \braket{S_1}-w\sigma_1 \\ \vdots \\ \braket{S_{2^n}}-w\sigma_{2^n}
    \end{bmatrix}}_{\bm{\beta_-}}\leq
   B\cdot\bm{p}
    \leq\underbrace{\begin{bmatrix}
    \braket{S_1}+w\sigma_1 \\ \vdots \\ \braket{S_{2^n}}+w\sigma_{2^n}
    \end{bmatrix}}_{\bm{\beta_+}},
\end{equation}
where $w\sigma_i$ with $w\geq0$ is the deviation to the mean value $\braket{S_i}$ represented in $\sigma_i$. To this end, an experimentally accessible constraint is formulated as $\bm{\beta_-}\leq B\bm{p}\leq \bm{\beta_+}$. In practice, instead of measuring all the stabilizers, which is impractical for a large quantum state, we can select a small subset of stabilizers so that the number of measurement does not scale exponentially to the number of qubits. Note that {$\langle\id^n\rangle=1$} must be set in Eq.~\ref{Eq:constraintInaeq} to ensure $\sum_kp_k=1$. {If we apply the scheme on the graph-diagonal states $\rho=\sum_k\lambda_k\ket{\psi_k}\bra{\psi_k}$ with $\bm{\lambda}=(\lambda_1,...,\lambda_{2^n})$ the spectrum of $\rho$, we have $\bm{p}=\bm{\lambda}$. Thus, $\bm{d} \vee (\wedge_{\bm{p} \in X}\bm{p})=\bm{\lambda}$ implies $l^c(\rho)=C_\text{RE}(\rho)$, which indicates that the estimated lower bound of coherence is \emph{tight} for graph-diagonal states.}

{We emphasize that relaxing the constraint to $\bm{\beta_-}\leq B\bm{p}\leq \bm{\beta_+}$ does not increasing the risk of overestimation of $l^c(\rho)$. Suppose that $X_1$ and $X_2$ are two feasible sets of probability distributions, and satisfy $X_1 \subseteq X_2$. $X_1$ and $X_2$ are restricted to $\bm{d} \prec \wedge_{\boldsymbol{p} \in X_1} \boldsymbol{p}$ and $\bm{d} \prec \wedge_{\boldsymbol{p} \in X_2} \boldsymbol{p} $, otherwise the result of Eq.~\ref{eq:MultpartiteEstimation} is 0. According to the definition of majorization meet, $\wedge_{\boldsymbol{p} \in X} \boldsymbol{p}$ is the ``largest'' distribution majorized by all probability distributions in $X$. Therefore, $\wedge_{\boldsymbol{p} \in X} \boldsymbol{p}$ becomes ``smaller''  when we enlarger the range of $X$, i.e, $\wedge_{\boldsymbol{p} \in X_1} \boldsymbol{p} \succ \wedge_{\boldsymbol{p} \in X_2} \boldsymbol{p}$ for $X_1 \subseteq X_2$. According to the strict Schur concavity of the Shannon entropy $S$, we obtain $S(\wedge_{\boldsymbol{p} \in X_1} \boldsymbol{p}) < S(\wedge_{\boldsymbol{p} \in X_2} \boldsymbol{p})$, which implies $S(\bm{d})- S(\wedge_{\boldsymbol{p} \in X_1} \boldsymbol{p}) > S(\bm{d})-S(\wedge_{\boldsymbol{p} \in X_2} \boldsymbol{p})$. Thus, we conclude that $l^c(\rho)$ decreases with enlarging the range of $X$. }

Similar constraints can also be formulated for multi-qubit states that do not obviously fit the stabilizer formalism. The stabilizing operators of such kind of \emph{n}-qubit states $\ket{\psi}$ could be determined by finding its unitary dynamics $U^\psi$ acting on $\ket{0}^{\otimes n}$, i.e., $\ket{\psi}=U^\psi\ket{0}^{\otimes n}$~\cite{nielsen_quantum_2010}. As $\ket{0}^{\otimes n}$ is stabilized by $S_i^{\ket{0}^{\otimes n}}=Z^{(i)}, \forall i\in\{1,2,...,n\}$, the stabilizing operator of $\ket{\psi}$ is $S_i^{\psi}=U^\psi Z^{(i)}{U^\psi}^{\dagger}$.

\subsection{Upper bound estimation}
\begin{theorem} Let $ \mathcal{M}(\bm{d})$ be a set of states with the same diagonal part $\bm{d}=(d_1,...,d_{2^n})$, then $\ket{\psi_d}=\sum_{i=1}^{2^n}\sqrt{d_i}\ket{i}$ is the maximally coherent state in $\mathcal{M}(\bm{d})$.
\end{theorem}

\begin{proof} It is equivalent to proof that $\ket{\psi_d}=\sum_{i=1}^{2^n}\sqrt{d_i}\ket{i}$ can be transformed into any $\rho \in \mathcal{M}(\bm{d})$ via incoherent operation. 

We first consider the case that there are \emph{no} non-zero elements in $\bm{d}$, i.e.,  $d_{i} \neq 0 $ for all $ i $. Let $\left\{\lambda_{\alpha},\left|\varphi_{\alpha}\right\rangle\right\}$ be the spectral decomposition of $\rho,$ i.e.,
\begin{eqnarray}
\rho =\sum_{\alpha} \lambda_{\alpha}\left|\varphi_{\alpha}\right\rangle\left\langle\varphi_{\alpha}\right| 
=\sum_{\alpha} \lambda_{\alpha} \sum_{i, j} c_{\alpha i} c_{\alpha j}^{*}|i\rangle\langle j|
\end{eqnarray}
where  $\left|\varphi_{\alpha}\right\rangle=\sum_{i=1}^{2^n}  c_{\alpha i}|i\rangle$ with $ \sum_{i=1}^{2^n} \left|c_{\alpha i}\right|^{2}=1$ for all $\alpha$.
Thus, $\rho$ form a set $\mathcal M(\bm{d})$ with $d_{i}=\sum_{\alpha} \lambda_{\alpha}\left|c_{\alpha i}\right|^{2}$. 
We define a completely positive and trace-preserving (CPTP) map $\Lambda_{1}(\cdot) = \sum_{\alpha} K _{\alpha}\cdot K_{\alpha}^{\dagger} $ associated with Kraus operators $K_{\alpha}=\sum_{i} \frac{\sqrt{\lambda_{\alpha}} c_{\alpha i}}{\sqrt{d_{i}}}|i\rangle\langle i|$, which transforms $\ket{\psi_d}\bra{\psi_d}$ into $\rho$ by
\begin{eqnarray}
\begin{aligned}
\Lambda_{1}(\ket{\psi_d}\bra{\psi_d})
&=\sum_{\alpha} K_{\alpha}|\psi_d\rangle\langle\psi_d| K_{\alpha}^{\dagger} \\&=\sum_{\alpha} \sum_{i, j, k ,l} \frac{\lambda_{\alpha} c_{\alpha i} c_{\alpha j}^{*}}{\sqrt{d_{i} d_{j}}} \sqrt{d_{k} d_{l}} \delta_{i k} \delta_{j l}|i\rangle\langle j| \\
&=\sum_{\alpha} \lambda_{\alpha} \sum_{i, j}  c_{\alpha i} c_{\alpha j}^{*}|i\rangle\langle j| \\
&=\rho.
\end{aligned}
\end{eqnarray} 
According to the definition of strictly incoherent operation \cite{yao_quantum_2015,de_vicente_genuine_2017}, $K_{\alpha}$ is a strictly incoherent operator and $\Lambda_{1}(\cdot)$ is a strictly incoherent operation.

For the case that there are $m$ ($m < 2^n$) non-zero elements in $\bm{d}$, we first transform $\ket{\psi_d}\bra{\psi_d}$ into a block diagonal matrix via via a permutation matrix $M$, i.e., $M\ket{\psi_d}\bra{\psi_d}M^{-1}=\ket{\psi_m}\bra{\psi_m} \oplus \bm{0}_{2^n-m}$. $\ket{\psi_m}= \sum_{i'=1}^{ m } \sqrt{d_{i{'}}}\ket{i'}$ is the vector in $m$-dimensional Hilbert space spanned by $\{\ket{i'}\}$, where $i^\prime$ is rearranged index of $m$ non-zero elements. $\bm{0}$ is a $(2^n - m)\times (2^n - m)$ all-0 matrix. 

Similarly, we define the CPTP map $\Lambda_{2}(\cdot) = \sum_{\beta} K _{\beta}\cdot K_{\beta}^{\dagger} $ associated with Kraus operators $K_{\beta}=\sum_{i'} \frac{\sqrt{\lambda_{\beta}} c_{\beta i'}}{\sqrt{d_{i'}}}|i'\rangle\langle i'|$ acting on $m$-dimensional Hilbert space. Thus, we can derive that 
\begin{eqnarray}
M^{-1}\Lambda_2(M\ket{\psi_d}\bra{\psi_d}M^{-1})M=\rho.
\end{eqnarray}
Note that any permutation matrix is strictly incoherent unitary \cite{liu_enhancing_2017}. For any two strictly incoherent operations $\Lambda_{\alpha }$ and $\Lambda_{\beta}$, the operation $\Lambda=\Lambda_{\alpha } \circ \Lambda_{\beta}$ is also a strictly incoherent operation \cite{liu_deterministic_2019}. Thus, $ \ket{\psi_d} =\sum_{i=1}^{2^n} \sqrt{d_{i}}|i\rangle$ can be transformed into any $\rho \in \mathcal{M}(\bm{d})$ via strictly incoherent operations $M^{-1} K_{\beta} M $.
\end{proof}
To upper bound the coherence of an \emph{n}-qubit state $\rho$, we measure it in the computational basis $\{\ket{i}\}$, which yields a distribution $\bm{d}=(d_1,...,d_{2^n})$. Based on the monogamy of the relative entropy of coherence~\cite{Baumgratz2014}, the coherence of $\rho$ is upper bounded by the coherence $u_c(\rho)$ of state $\ket{\psi_d}=\sum_{i=1}^{2^n}\sqrt{d_i}\ket{i}$, i.e.,
\begin{equation}\label{Eq:upperbound}
    C_{\text{RE}}(\rho)\leq u_c(\rho)=C_{\text{RE}}(\ket{\psi_d}\bra{\psi_d}).
\end{equation}
Note that this lower bound is tight for pure states, and is capable for various coherence measures.  

\begin{figure*}[htbp]
\includegraphics[width=1.8\columnwidth]{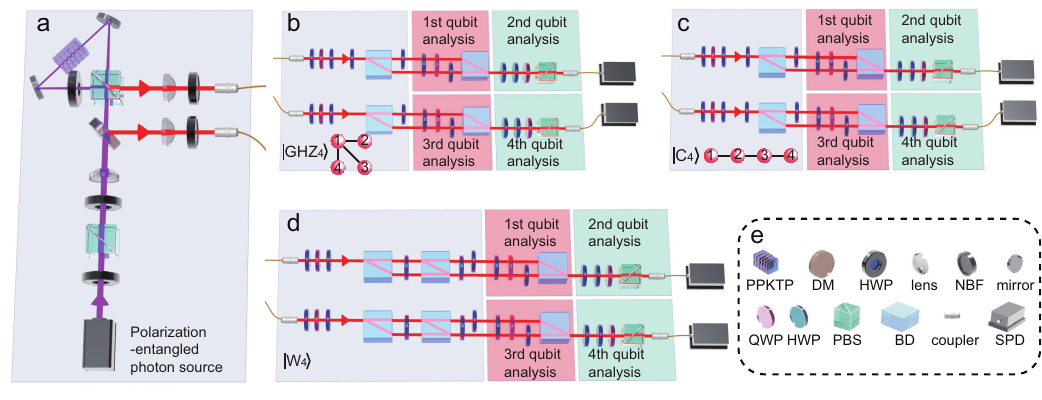} 
\caption{
Schematic drawing of experimental setup. a, the setup to generate polarization-entangled photon pair. b-d, the setups to generate $\ket{\text{GHZ}_4}$, $\ket{\text{C}_4}$ and $\ket{\text{W}_4}$ respectively. e, Symbols used in b, c, and d: periodically poled potassium titanyl phosphate (PPKTP), dichroic mirror (DM), narrow-band filter (NBF), half-wave plate (HWP), quarter-wave plate (QWP), polarization beam splitter (PBS), beam displacer (BD) and single-photon detector (SPD). More experimental details can be found in~\cref{Sec:Expdetails}.}
\label{Fig:ExpSetup}
\end{figure*}

\section{Experimental demonstration}
Next, we demonstrate the capability of our scheme by estimating the coherence of several typical multi-qubit states. We firstly generate photon pairs by a periodically poled potassium titanyl phosphate (PPKTP) crystal in a Sagnac interferometer~\cite{Kim2006_entangledphoton}, which is bidirectionally pumped by an ultraviolet (UV) laser diode with central wavelength at 405~nm (as shown in Fig.~\ref{Fig:ExpSetup}a). The two photons are entangled in the polarization degree of freedom (DOF), i.e., $\ket{\Psi^{+}_{ab}}=(\ket{H_aV_b}+\ket{V_aH_b})/\sqrt{2}$ with $H$ the horizontal polarization and $V$ the vertical polarization. We extend photon to its path DOF by beam displacer (BD), which transmits vertical polarization and deviate horizontal polarization, i.e., $\ket{H}\to\ket{H}\ket{h}$ and $\ket{V}\to\ket{V}\ket{v}$ with $h$ and $v$ the path DOF~\cite{Kai2007, Vallone2008, Weibo2010}. The qubit is encoded polarization DOF as $\ket{H(V)}\to\ket{0(1)}$, and path DOF as $\ket{h(v)}\to\ket{0(1)}$. In our experiment, we denote the qubits encoded in polarization DOF as $1$ and $3$, while the qubits encoded in path DOF as $2$ and $4$. As shown in Fig.~\ref{Fig:ExpSetup}b-d, with different experimental setup configurations, we can generate various 4-qubit states, including $\ket{\text{GHZ}_4}=(\ket{0000}+\ket{1111})/\sqrt{2}$, $\ket{\text{C}_4}=(\ket{0000}+\ket{0011}+\ket{1100}-\ket{1111})/2$ and $\ket{\text{W}_4}=(\ket{0001}+\ket{0010}+\ket{0100}+\ket{1000})/2$. Moreover, $\ket{\text{GHZ}}_3=(\ket{000}+\ket{111})/\sqrt{2}$ and $\ket{\text{W}_3}=(\ket{100}+\ket{010}+\ket{001})/\sqrt{3}$ can be obtained by extending one photon to polarization and path DOF while keeping the other in polarization DOF (See \cref{Sec:Expdetails} for more details).

\begin{figure}[h!tbp]
\includegraphics[width=1\columnwidth]{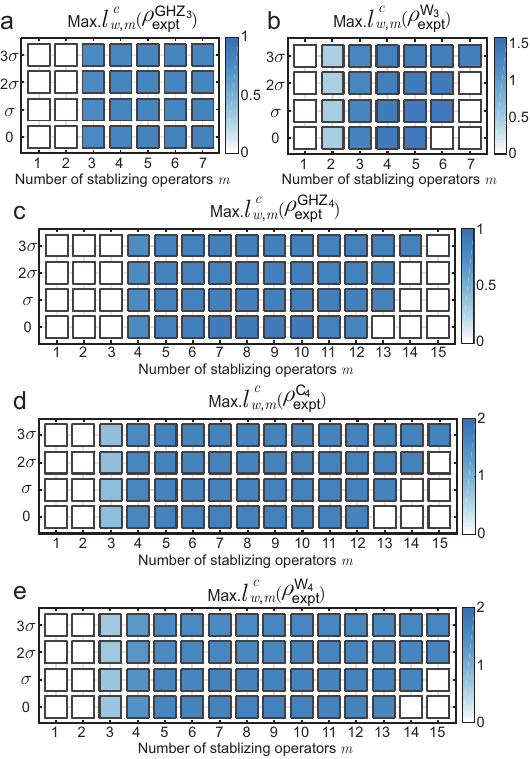} 
\caption{Experimental results of $l^c_{\omega, m}(\rho_{\text{expt}}^{\psi})$. The color bar indicates the range of $l^c_{\omega, m}(\rho_{\text{expt}}^{\psi})$ from 0 to values of ideal $\ket{\psi}$, i.e., 1 for $\ket{\text{GHZ}_3}$ and $\ket{\text{GHZ}_4}$, 1.585 for $\ket{\text{W}_3}$ and 2 for $\ket{\text{C}_4}$ and $\ket{\text{W}_4}$. }
\label{Fig:data1}
\end{figure}

\begin{figure*}[ht!]
\includegraphics[width=1.8\columnwidth]{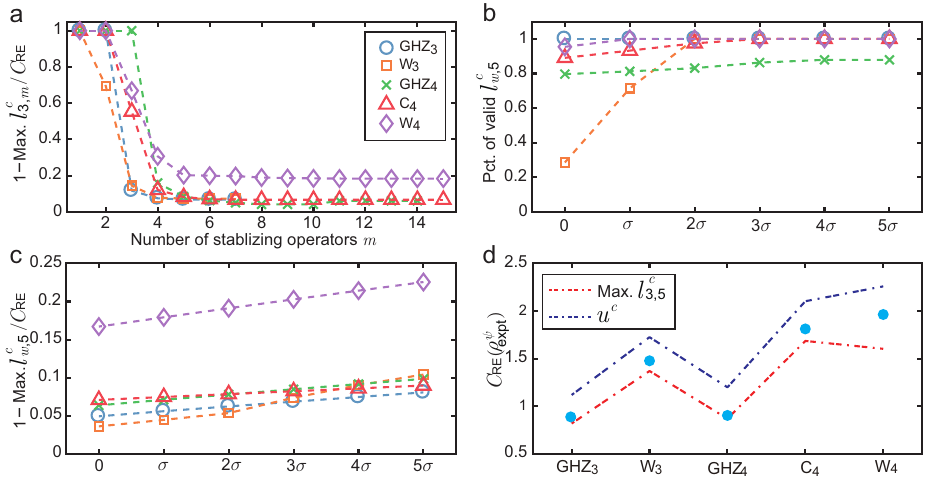} 
\caption{a, the normalized distance between maximal $l^c_{3, m}(\rho_{\text{expt}}^{\psi})$ and $C_{\text{RE}}(\rho_{\text{expt}}^{\psi})$. b, the percentage of valid $l^c_{\omega, 5}$ obtained over all $\binom{2^n-1}{5}$ subsets. c, the normalized distance between maximal $l^c_{\omega, 5}(\rho_{\text{expt}}^{\psi})$ and $C_{\text{RE}}(\rho_{\text{expt}}^{\psi})$ of setting $\omega$ from 0 to 5. d, the $u^c$ (blue dash line) for $\rho_{\text{expt}}^{\text{GHZ}_3}$, $\rho_{\text{expt}}^{\text{W}_3}$,  $\rho_{\text{expt}}^{\text{GHZ}_4}$, $\rho_{\text{expt}}^{\text{C}_4}$ and $\rho_{\text{expt}}^{\text{W}_4}$ are $1.117\pm0.003$, $1.725\pm0.004$, $1.198\pm0.005$, $2.103\pm0.004$, $1.725\pm0.004$ and $2.259\pm0.006$, respectively. The corresponding maximal $l^c_(\omega, m)$ (red dash line) are the results of $\omega=3$ and $m=5$. The dots represent $C_{\text{RE}}(\rho_{\text{expt}}^{\psi})$, and their error bars are too small compared to the marker size (See \cref{Sec:Expdetails} for the values of $C_{\text{RE}}(\rho_{\text{expt}}^{\psi})$.)}
\label{Fig:data2}
\end{figure*}

For each experimentally generated state $\rho_{\text{expt}}^{\psi}$, we measure the expected values of its stabilizers $\braket{S_i}$ associated with the corresponding statistical errors $\sigma_i$. We refer to {\cref{Sec:Stablizer} for details of stabilizing operators of $\ket{\psi}$ and the corresponding graph-diagonal basis. The measured expected values of stabilizing operators are presented in \cref{Sec:Expdetails}. With the measured $\braket{S_i}$ and $\sigma_i$, we construct the} constraint Eq.~\ref{Eq:constraintInaeq}.  {Thus, we can calculate $l^c_{\omega, m}(\rho_{\text{expt}}^{\psi})$ via solving } Eq.~\ref{eq:MultpartiteEstimation}, where $\omega$ represents the setting of deviations and $m$ is the number of $\langle S_i\rangle$ we {construct the constraint}. We set $\omega$ as non-negative integers from 0 to 3, and $m$ from 1 to $2^n-1$ as {$\braket{\id^n}=1$} must be set in the constraint. In our calculation, we treat the case of no solution as $l_{\omega, m}^c(\rho_{\text{expt}}^{\psi}) = 0$, and the case of $l_{\omega, m}^c(\rho_{\text{expt}}^{\psi})>0$ as valid solution.  For a fixed $m$, there are $\binom{2^n-1}{m}$ subsets of $\{S_i^\psi\}$, and the maximum $l^c_{\omega, m}(\rho_{\text{expt}}^{\psi})$ is shown in Fig.~\ref{Fig:data1}. We observe that valid $l^c_{\omega, m}(\rho_{\text{expt}}^{\psi})$ can not be obtained with $m\leq 3$ stabilizers for $\rho_{\text{expt}}^{\text{GHZ}_4}$, $m\leq 1$ stabilizers for $\rho_{\text{expt}}^{\text{W}_3}$, and $m\leq 2$ stabilizers for other three states. With the increasing of $m$, the maximum $l^c_{\omega, m}(\rho_{\text{expt}}^{\psi})$ increases accordingly. When $m$ gets close to $2^n-1$, there exists situations that we can not obtain $l^c_{\omega, m}(\rho_{\text{expt}}^{\psi})$ for smaller $w$ {(the values are 0 at the right lower corner of each figure in \cref{Fig:data1})}. {As aforementioned, }this is caused by the experimental imperfections, such as slight misalignment of optical elements during data collection, which introduces small variation of prepared $\rho_{\text{expt}}^{\psi}$. The issue is improved by extending the range of $\braket{S_i^\psi}$, i.e., increasing $\omega$. As shown in \cref{Fig:data1}b-\cref{Fig:data1}e, most $l^c_{\omega, m}(\rho_{\text{expt}}^{\psi})$ has valid solutions for large $m$ by setting $\omega=3$. Moreover, the accuracy of estimated $l^c_{\omega, m}(\rho_{\text{expt}}^{\psi})$ increases along with $m$ as well. We investigate this by calculating the normalized distance between $l^c_{\omega, m}(\rho_{\text{expt}}^{\psi})$ and $C_{\text{RE}}(\rho_{\text{expt}}^{\psi})$, i.e., $1-l^c_{\omega, m}(\rho_{\text{expt}}^{\psi})/C_{\text{RE}}(\rho_{\text{expt}}^{\psi})$. $C_{\text{RE}}(\rho_{\text{expt}}^{\psi})$ is calculated with Eq.~\ref{Eq:RE} by reconstructing $\rho_{\text{expt}}^{\psi}$ via quantum state tomographic technology (See \cref{Sec:Expdetails} for the reconstructed $\rho_{\text{expt}}^{\psi}$). The distances between maximal $l^c_{3, m}(\rho_{\text{expt}}^{\psi})$ and $C_{\text{RE}}(\rho_{\text{expt}}^{\psi})$ are shown in Fig.~\ref{Fig:data2}a, from which we observe that the distance drops down quickly with increasing of $m$ and tends to converge at $m=5$. 

As aforementioned, the choice of selecting $m$ stabilizers from $\{S_i^\psi\}$ is not unique except $m=2^n-1$. An important property is the successful probability of obtaining valid $l^c_{\omega, m}$ by randomly selecting $m$ stabilizers. We show the percentage of valid $l^c_{\omega, 5}(\rho_{\text{expt}}^{\psi})$ ($l^c_{\omega, 5}(\rho_{\text{expt}}^{\psi})>0$) for $m=5$ in Fig.~\ref{Fig:data2}b. By increasing $\omega$, the probability of getting valid $l^c_{\omega, 5}(\rho_{\text{expt}}^{\psi})$ is enhanced, especially for $\rho_{\text{expt}}^{\text{W}_3}$. However, one cannot increasing $\omega$ arbitrarily. A larger $\omega$ represents a smaller probability we could obtain $\braket{S_i}$ in range $[\braket{S_i}+(w-1)\sigma_i,\braket{S_i}+w\sigma_i]$ as well as $[\braket{S_i}-w\sigma_i,\braket{S_i}-(w-1)\sigma_i]$, which is less than 0.3\% for $\omega=3$. This is also reflected by normalized distance of maximal $l^c_{\omega, 5}(\rho_{\text{expt}}^{\psi})$ shown in Fig.~\ref{Fig:data2}c, which indicates the inaccuracy of $l^c_{\omega, 5}(\rho_{\text{expt}}^{\psi})$ increases when we extend the range of $\braket{S_i}$. {Also, the results in Fig.~\ref{Fig:data2}c agree with our claim that relaxing the constraint decreases the estimated value of $l^c(\rho)$. We conclude that set $\omega=3$ is reasonable in experiment, under which} we observe the probability of getting valid $l^c_{\omega, 5}(\rho_{\text{expt}}^{\psi})$ is 100\% for $\rho_{\text{expt}}^{\text{GHZ}_3}$,  $\rho_{\text{expt}}^{\text{W}_3}$,  $\rho_{\text{expt}}^{\text{C}_4}$ and  $\rho_{\text{expt}}^{\text{W}_4}$, and that of 86\% for  $\rho_{\text{expt}}^{\text{GHZ}_4}$. {Note that the estimated $l^c(\rho)$ of $\rho_{\text{expt}}^{\text{GHZ}_3}$,  $\rho_{\text{expt}}^{\text{W}_3}$,  $\rho_{\text{expt}}^{\text{C}_4}$ and $\rho_{\text{expt}}^{\text{GHZ}_4}$ is slightly more accurate than that of $\rho_{\text{expt}}^{\text{W}_4}$ as shown in \cref{Fig:data2}a and \cref{Fig:data2}c. The main reason is that the fidelity of prepared $\rho_{\text{expt}}^{\text{W}_4}$ is slightly lower than that of other four states as shown in \cref{Sec:Expdetails}. The lower fidelity implies a larger distance between prepared state and target state, which indicates $\bm{\lambda} \succ \bm{d} \vee(\wedge_{\boldsymbol{p} \in X} \boldsymbol{p})$ so that $C_\text{RE}(\rho)=S(\bm{d})- S(\bm{\lambda}) > l^c(\rho)=S(\bm{d})-S(\bm{d} \vee(\wedge_{\boldsymbol{p} \in X} \boldsymbol{p}))$. However, it does not indicate the lower fidelity always leads to the bigger gap between $C_\text{RE}(\rho)$ and $l^c(\rho)$. If the prepared state is still a graph-diagonal state after considering the experimental imperfections, the estimated lower bound on such state is tight as well, i.e., $l^c(\rho)=C_\text{RE}(\rho)$. In \cref{Sec:Noise}, we analyze our experimental imperfections and show how it affect the prepared states.}

Finally, we estimate the upper bound $u^c$ of $\rho_{\text{expt}}^{\psi}$ by measuring the probability distribution $\bm{d}^\psi_\text{expt}$ on basis of $Z^{\otimes n}$. The probability distribution $\bm{d}^\psi_\text{expt}$ are shown in \cref{Sec:Expdetails}, by which we can calculate $u^c(\rho_{\text{expt}}^{\psi})$ according to Eq.~\ref{Eq:upperbound}. The results of $u^c(\rho_{\text{expt}}^{\psi})$ are shown with blue dash line in Fig.~\ref{Fig:data2}d. For all the five states, we observe that $C_{\text{RE}}(\rho_\text{expt}^\psi$) lies within the range bounded by $l^c_{3, 5}(\rho_{\text{expt}}^{\psi})$ and $u^c(\rho_{\text{expt}}^{\psi})$.

\section{Conclusion}
To conclude, we introduce an efficient and experimentally friendly estimation method for detecting coherence of multipartite states. We demonstrate that the coherence with high accuracy as well as high successful probability can be efficiently estimated with a few measurements for various multi-qubit states. The procedure to obtain the lower bound is based on few measurements of the stabilizing operators, similar to multipartite entanglement detection~\cite{Toth2005} and multipartite Bell inequalities~\cite{zhao2020constructing}. It thus indicates that coherence and other resources can be inferred by the same set of measurements, which can further benefit our understanding of the connection between coherence and other resources~\cite{streltsov_measuring_2015,Chitambar_coherence-entanglement2016,Girolami2017, killoran_converting_2016}. 

{There are several} open follow-up problems. {The scheme to detect the lower bound of multipartite coherence is efficient and tight for graph-diagonal states as well as special types of quantum states that admit efficient classical representation (such as the W state). Whether our scheme will work for more general multi-qubit quantum state is an interesting future work.} Besides, one may concern whether our scheme can be generalized to high-dimensional cases. For the multi-qudit stabilizer state, its stabilizing operators constructed by generalized Pauli group are generally non-Hermitian~\cite{Gottesman1999,Hostens2005,Gheorghiu2014}, which can not be observed directly in experiment.  Recent studies indicate that expected values of non-Hermitian operators could be measured via weak measurements~\cite{Pati2015, Nirala2019}. However, the definite answer may require rather sophisticated analysis. {Finally}, it is worth noting that a fidelity-based method was recently proposed to detect the lower bound of multipartite coherence via the convex roof construction~\cite{DaiPhysRevApplied2020}. Thus, another open follow-up question is whether our scheme can be further generalized to other coherence measures while maintaining their appealing feature of efficiency and simplicity.  

\begin{acknowledgments}
We are grateful to anonymous referee for providing very useful comments on an earlier version of this manuscript. This work is supported by the National Natural Science Foundation of China (No. 11974213), National Key R \& D Program of China (No. 2019YFA0308200) and Shandong Provincial Natural Science Foundation (No. ZR2019MA001).
\end{acknowledgments}

\appendix

\section{Stabilizing operators and graph-diagonal basis}
\label{Sec:Stablizer}
\subsection{GHZ state}
For $n$-qubit GHZ states, the generators of $\mathcal{S}^{\text{GHZ}}$ are 
\begin{equation}\label{Eq:GHZGenerator}
\begin{split}
    &S_1^{\text{GHZ}_n}=\prod_{i=1}^{n}X^{(i)},\\
    &S_i^{\text{GHZ}_n}=Z^{(i-1)}Z^{(i)}\quad\text{for}\quad i=2,3,...,n, 
\end{split}
\end{equation}
and denoted as $\mathcal{S}^{\text{GHZ}_n}=\langle S_1^{\text{GHZ}_n}, S_2^{(\text{GHZ}_n)}, ..., S_n^{\text{GHZ}_n}\rangle$. For $\ket{\text{GHZ}_3}$, the generators are $S_1^{\text{GHZ}_3}=XXX$, $S_2^{\text{GHZ}_3}=ZZ\mathbb{I}$ and $S_3^{\text{GHZ}_3}=\mathbb{I}ZZ$. Then, all stabilizing operators can be obtained by multiplying them with each other, i.e.,
\begin{equation}
    \begin{split}
        &S_4^{\text{GHZ}_3}=S_1^{\text{GHZ}_3}S_2^{\text{GHZ}_3}=-YYX,\\
        &S_5^{\text{GHZ}_3}=S_1^{\text{GHZ}_3}S_3^{\text{GHZ}_3}=-XYY,\\
        &S_6^{\text{GHZ}_3}=S_2^{\text{GHZ}_3}S_3^{\text{GHZ}_3}=Z\mathbb{I}Z,\\
        &S_7^{\text{GHZ}_3}=S_1^{\text{GHZ}_3}S_2^{\text{GHZ}_3}S_3^{\text{GHZ}_3}=-YXY,\\
        &S_8^{\text{GHZ}_3}=\mathbb{III}.\\
    \end{split}
\end{equation}
{
The $\text{GHZ}_3$-diagonal basis can be obtained from the computational basis $\ket{k_1k_2k_3}$ by acting a 3-qubit unitary operation $U^{\text{GHZ}_3}=(\text{CNOT}_{12}\otimes \id_3)\cdot(\otimes\text{CNOT}_{13}\otimes\id_2)\cdot(\otimes H_1\id_2\otimes\id_3)$, where $H$ is the Hadamard operation and
\begin{equation}
\text{CNOT}=\begin{pmatrix}1&0&0&0\\0&1&0&0\\0&0&0&1\\0&0&1&0\end{pmatrix}
\end{equation}
is the unitary matrix of controlled-not (CNOT) operation. Eight $\text{GHZ}_3$-diagonal bases are determined by $\ket{\psi_i^{\text{GHZ}_3}}=U^{\text{GHZ}_3}\ket{k_1k_2k_3}$ and shown in \cref{tab:GHZ3bases}.
\begin{table}[h]
\vspace{0.2cm}
\begin{tabular}{|c|c|c|}
\hline
$k_1k_2k_3$     & $\ket{\psi_i^{\text{GHZ}_3}}$     & $\tiny(\langle S_1^{\text{GHZ}_3}\rangle, \langle S_2^{\text{GHZ}_3}\rangle, \langle S_3^{\text{GHZ}_3}\rangle)$  \\ \hline
000  & $(\ket{000}+\ket{111})/\sqrt{2}$ & $(+1, +1, +1)$ \\ \hline
001  & $(\ket{000}-\ket{111})/\sqrt{2}$ & $(+1, +1, -1)$ \\ \hline
010  & $(\ket{001}+\ket{110})/\sqrt{2}$ & $(-1, +1, +1)$ \\ \hline
011  & $(\ket{001}-\ket{110})/\sqrt{2}$ & $(-1, +1, -1)$ \\ \hline
100  & $(\ket{011}+\ket{100})/\sqrt{2}$ & $(+1, -1, +1)$ \\ \hline
101  & $(\ket{010}+\ket{101})/\sqrt{2}$ & $(+1, -1, -1)$ \\ \hline
110  & $(\ket{100}-\ket{011})/\sqrt{2}$ & $(-1, -1, +1)$ \\ \hline
111  & $(\ket{101}-\ket{010})/\sqrt{2}$ & $(-1, -1, -1)$ \\ \hline
\end{tabular}
\caption{The $\text{GHZ}_3$-diagonal bases and the corresponding expected values of generators $(\langle S_1^{\text{GHZ}_3}\rangle, \langle S_2^{\text{GHZ}_3}\rangle, \langle S_3^{\text{GHZ}_3}\rangle)$. }
\label{tab:GHZ3bases}
\end{table}
}

The generators of $\ket{\text{GHZ}_4}$ are $S_1^{\text{GHZ}_4}=XXXX$, $S_2^{\text{GHZ}_4}=ZZ\mathbb{II}$, $S_3^{\text{GHZ}_4}=\mathbb{I}ZZ\mathbb{I}$ and $S_4^{\text{GHZ}_4}=\mathbb{II}ZZ$. Other stabilizing operators are
\begin{equation}
    \begin{split}
        &S_5^{\text{GHZ}_4}=-YYXX, S_6^{\text{GHZ}_4}=-XYYX,\\
        &S_7^{\text{GHZ}_4}=-XXYY, S_8^{\text{GHZ}_4}=Z\mathbb{I}Z\mathbb{I},\\
        &S_9^{\text{GHZ}_4}=ZZZZ, S_{10}^{\text{GHZ}_4}=\mathbb{I}Z\mathbb{I}Z,\\
        &S_{11}^{\text{GHZ}_4}=-YXYX, S_{12}^{\text{GHZ}_4}=YYYY,\\
        &S_{13}^{\text{GHZ}_4}=-XYXY, S_{14}^{\text{GHZ}_4}=Z\mathbb{II}Z,\\
        &S_{15}^{\text{GHZ}_4}=-YXXY, S_{16}^{\text{GHZ}_4}=\mathbb{IIII}.\\
    \end{split}
\end{equation}
{The $\text{GHZ}_4$-diagonal basis can be obtained from the computational basis $\ket{k_1k_2k_3k_4}$ by acting a 4-qubit unitary operation $U^{\text{GHZ}_4}=(\id_1\otimes H_2\otimes H_3 \otimes H_4)\cdot(\text{CZ}_{14})\cdot(\text{CZ}_{13})\cdot(\text{CZ}_{12})\cdot(H_1\otimes H_2\otimes H_3\otimes H_4)$, where
\begin{equation}
\text{CZ}=\begin{pmatrix}1&0&0&0\\0&1&0&0\\0&0&1&0\\0&0&0&-1\end{pmatrix}
\end{equation}
is the unitary matrix of controlled-Z (CZ) operation.
Sixteen $\text{GHZ}_4$-diagonal bases are determined by $\ket{\psi_i^{\text{GHZ}_4}}=U^{\text{GHZ}_4}\ket{k_1k_2k_3k_4}$ and shown in \cref{tab:GHZ4bases}.
\begin{table}[h]
	\vspace{0.2cm}
	\begin{tabular}{|c|c|c|}
		\hline
		$k_1k_2k_3k_4$     & $\ket{\psi_i^{\text{GHZ}_4}}$     & \tiny($\braket{S_1^{\text{GHZ}_4}}$, $\braket{S_2^{\text{GHZ}_4}}$, $\braket{S_3^{\text{GHZ}_4}}$, $\braket{S_4^{\text{GHZ}_4}}$)  \\ \hline
		0000  & $(\ket{0000}+\ket{1111})/\sqrt{2}$ & $(+1, +1, +1 ,+1)$ \\ \hline
		0001  & $(\ket{0001}+\ket{1110})/\sqrt{2}$ & $(+1, +1 ,+1, -1)$ \\ \hline
		0010  & $(\ket{0010}+\ket{1101})/\sqrt{2}$ & $(+1, +1, -1 ,-1)$ \\ \hline
		0011  & $(\ket{0011}+\ket{1100})/\sqrt{2}$ & $(+1, +1, -1 ,+1)$ \\ \hline
		0100  & $(\ket{0100}+\ket{1011})/\sqrt{2}$ & $(+1, -1, -1,+1)$ \\ \hline
		0101  & $(\ket{0101}+\ket{1010})/\sqrt{2}$ & $(+1, -1, -1,-1)$ \\ \hline
		0110  & $(\ket{0110}+\ket{1001})/\sqrt{2}$ & $(+1, -1, +1,-1)$ \\ \hline
		0111  & $(\ket{0111}+\ket{1000})/\sqrt{2}$ & $(+1, -1, +1,+1)$ \\ \hline
		1000  & $(\ket{0000}-\ket{1111})/\sqrt{2}$ & $(-1, +1, +1,+1)$ \\ \hline
		1001  & $(\ket{0001}-\ket{1110})/\sqrt{2}$ & $(-1, +1, +1,-1)$ \\ \hline
		1010  & $(\ket{0010}-\ket{1101})/\sqrt{2}$ & $(-1, +1, -1,-1)$ \\ \hline
		1011  & $(\ket{0011}-\ket{1100})/\sqrt{2}$ & $(-1, +1, -1,+1)$ \\ \hline
		1100  & $(\ket{0100}-\ket{1011})/\sqrt{2}$ & $(-1, -1, -1,+1)$ \\ \hline
		1101  & $(\ket{0101}-\ket{1010})/\sqrt{2}$ & $(-1, -1, -1,-1)$ \\ \hline
		1110  & $(\ket{0110}-\ket{1001})/\sqrt{2}$ & $(-1, -1, +1,-1)$ \\ \hline
		1111  & $(\ket{0111}-\ket{1000})/\sqrt{2}$ & $(-1, -1, +1,+1)$ \\ \hline		
	\end{tabular}
	\caption{The $\text{GHZ}_4$-diagonal bases and the corresponding expected values of generators $(\langle S_1^{\text{GHZ}_4}\rangle, \langle S_2^{\text{GHZ}_4}\rangle, \langle S_3^{\text{GHZ}_4}\rangle, \langle S_4^{\text{GHZ}_4}\rangle)$. }
	\label{tab:GHZ4bases}
\end{table}
}

\subsection{Cluster state}
The generators of a $n$-qubit cluster state $\ket{\widetilde{\text{C}}_n}$ are
\begin{equation}\label{Eq:ClusterGenerator}
    \begin{split}
        &S_1^{\widetilde{\text{C}}_n}=X^{(1)}Z^{(2)},\\
        &S_i^{\widetilde{\text{C}}_n}=Z^{(i-1)}X^{(i)}Z^{(i-1)}\quad\text{for}\quad i=2,3,...,n-1,\\
        &S_n^{\widetilde{\text{C}}_n}=Z^{(n-1)}X^{(n)}.
    \end{split}
\end{equation}
The generators of 4-qubit linear graph are $S_1^{\widetilde{\text{C}}_4}=XZ\mathbb{II}$, $S_2^{\widetilde{\text{C}}_4}=ZXZ\mathbb{I}$, $S_3^{\widetilde{\text{C}}_4}=\mathbb{I}ZXZ$ and $S_4^{\widetilde{\text{C}}_4}=\mathbb{II}ZX$, and the corresponding state is $\ket{\widetilde{\text{C}}_4}=(\ket{+00+}+\ket{+01-}+\ket{-10+}-\ket{-11-})/2$. Note that $\ket{\widetilde{\text{C}}_4}$ can be transformed to the common representation $\ket{\text{C}_4}=(\ket{0000}+\ket{0011}+\ket{1100}-\ket{1111})/2$ by local unitary $H\mathbb{II}H$, i.e., $\ket{\text{C}_4}=H\mathbb{II}H\ket{\widetilde{\text{C}}_4}$. Accordingly, the generators of $\ket{\text{C}_4}$ are transformed to $ZZ\mathbb{II}$, $XXZ\mathbb{I}$, $\mathbb{I}ZXX$ and $\mathbb{II}ZZ$. Other stabilizing operators are
\begin{equation}
    \begin{split}
        &S_5^{\text{C}_4}=-YYZ\mathbb{I}, S_6^{\text{C}_4}=Z\mathbb{I}XX,\\
        &S_7^{\text{C}_4}=ZZZZ, S_8^{\text{C}_4}=XYYX,\\
        &S_9^{\text{C}_4}=XX\mathbb{I}Z, S_{10}^{\text{C}_4}=-\mathbb{I}ZYY,\\
        &S_{11}^{\text{C}_4}=YXYX, S_{12}^{\text{C}_4}=-YY\mathbb{I}Z,\\
        &S_{13}^{\text{C}_4}=-Z\mathbb{I}YY, S_{14}^{\text{C}_4}=XYXY,\\
        &S_{15}^{\text{C}_4}=YXXY, S_{16}^{\text{C}_4}=\mathbb{IIII}.\\ 
    \end{split}
\end{equation}
{
	The $\text{Cluster}$-diagonal basis can be obtained from the computational basis $\ket{k_1k_2k_3k_4}$ by acting a 4-qubit unitary operation $U^{\text{C}_4}=(\id_1\otimes\text{CZ}_{23}\otimes \id_4)\cdot(\text{CNOT}_{12}\otimes\text{CNOT}_{34})\cdot(H_1\otimes\id_2\otimes H_3\otimes\id_4)$. Then, the $\text{Cluster}$-diagonal bases are determined by $\ket{\psi_i^{\text{C}_4}}=U^{\text{C}_4}\ket{k_1k_2k_3k_4}$ and shown in \cref{tab:C4bases}.
	\begin{table}[h]
		\vspace{0.2cm}
	\begin{tabular}{|c|c|c|}
	\hline
	$k_1k_2k_3k_4$     & $\ket{\psi_i^{\text{C}_4}}$     &\tiny($\braket{S_1^{\text{C}_4}}$, $\braket{S_2^{\text{C}_4}}$, $\braket{S_3^{\text{C}_4}}$, $\braket{S_4^{\text{C}_4}}$)  \\ \hline
	0000  & $\frac{\ket{0000}+\ket{0011}+\ket{1100}-\ket{1111}}{2}$ & $(+1, +1, +1 ,+1)$ \\ \hline
	0001  & $\frac{\ket{0001}+\ket{0010}+\ket{1101}-\ket{1110}}{2}$ & $(+1, +1 ,+1, -1)$ \\ \hline
	0010  & $\frac{\ket{0000}-\ket{0011}+\ket{1100}+\ket{1111}}{2}$ & $(+1, +1, -1 ,+1)$ \\ \hline
	0011  & $\frac{\ket{0001}-\ket{0010}+\ket{1101}+\ket{1110}}{2}$ & $(+1, +1, -1 ,-1)$ \\ \hline
	0100  & $\frac{\ket{0100}-\ket{0111}+\ket{1000}+\ket{1011}}{2}$ & $(-1, +1, +1,+1)$ \\ \hline
	0101  & $\frac{\ket{0101}-\ket{0110}+\ket{1001}+\ket{1010}}{2}$ & $(-1, +1, +1,-1)$ \\ \hline
	0110  & $\frac{\ket{0100}+\ket{0111}+\ket{1000}-\ket{1011}}{2}$ & $(-1, +1, -1,+1)$ \\ \hline
	0111  & $\frac{\ket{0101}+\ket{0110}+\ket{1001}-\ket{1010}}{2}$ & $(-1, +1, -1,-1)$ \\ \hline
	1000  & $\frac{\ket{0000}+\ket{0011}-\ket{1100}+\ket{1111}}{2}$ & $(+1, -1, +1,+1)$ \\ \hline
	1001  & $\frac{\ket{0001}+\ket{0010}-\ket{1101}+\ket{1110}}{2}$ & $(+1, -1, +1,-1)$ \\ \hline
	1010  & $\frac{\ket{0000}-\ket{0011}-\ket{1100}-\ket{1111}}{2}$ & $(+1, -1, -1,+1)$ \\ \hline
	1011  & $\frac{\ket{0001}-\ket{0010}-\ket{1101}-\ket{1110}}{2}$ & $(+1, -1, -1,-1)$ \\ \hline
	1100  & $\frac{\ket{0100}-\ket{0111}-\ket{1000}-\ket{1011}}{2}$ & $(-1, -1, +1,+1)$ \\ \hline
	1101  & $\frac{\ket{0101}-\ket{0110}-\ket{1001}-\ket{1010}}{2}$ & $(-1, -1, +1,-1)$ \\ \hline
	1110  & $\frac{\ket{0100}+\ket{0111}-\ket{1000}+\ket{1011}}{2}$ & $(-1, -1, -1,+1)$ \\ \hline
	1111  & $\frac{\ket{0101}+\ket{0110}-\ket{1001}+\ket{1010}}{2}$ & $(-1, -1, -1,-1)$ \\ \hline		
\end{tabular}
		\caption{The $\text{Cluster}$-diagonal bases and the corresponding expected values of generators $(\langle S_1^{\text{C}_4}\rangle, \langle S_2^{\text{C}_4}\rangle, \langle S_3^{\text{C}_4}\rangle, \langle S_4^{\text{C}_4}\rangle)$. }
		\label{tab:C4bases}
	\end{table}
}

\subsection{W state}
$\ket{\text{W}_3}$ can be transformed from $\ket{000}$ by unitary $U^{\text{W}_3}=(XZ\mathbb{I}+\mathbb{I}XZ+Z\mathbb{I}X)/\sqrt{3}$, i.e., $\ket{\text{W}_3}=U^{\text{W}_3}\ket{000}$\cite{Toth2005}. Thus the generators of $\ket{\text{W}_3}$ is derived by 
\begin{equation}
    \begin{split}
        &S_{1}^{\text{W}_3}=U^{\text{W}_3}Z\mathbb{II}{U^{\text{W}_3}}^\dagger=\frac{1}{3}(Z\mathbb{II}+2YYZ+2XZX), \\
        &S_{2}^{\text{W}_3}=U^{\text{W}_3}\mathbb{I}Z\mathbb{I}{U^{\text{W}_3}}^\dagger=\frac{1}{3}(\mathbb{I}Z\mathbb{I}+2ZYY+2XXZ), \\
        &S_{3}^{\text{W}_3}=U^{\text{W}_3}\mathbb{II}Z{U^{\text{W}_3}}^\dagger=\frac{1}{3}(\mathbb{II}Z+2YZY+2ZXX).
    \end{split}
\end{equation}
Other stabilizing operators are
\begin{equation}
    \begin{split}
        &S_4^{\text{W}_3}=\frac{1}{3}(2X\mathbb{I}X+2\mathbb{I}YY-ZZ\mathbb{I}),\\
        &S_5^{\text{W}_3}=\frac{1}{3}(2\mathbb{I}XX+2YY\mathbb{I}-Z\mathbb{I}Z),\\
        &S_6^{\text{W}_3}=\frac{1}{3}(2XX\mathbb{I}+2Y\mathbb{I}Y-\mathbb{I}ZZ),\\
        &S_7^{\text{W}_3}=-ZZZ, S_8^{\text{W}_3}=\mathbb{III}.\\
    \end{split}
\end{equation}
{The $\text{W}_3$-diagonal bases are shown in \cref{tab:W3bases}.
\begin{table}[h]
\vspace{0.2cm}
\begin{tabular}{|c|c|c|}
\hline
$k_1k_2k_3$     & $\ket{\psi_i^{\text{W}_3}}$     & $(\braket{S_{1}^{\text{W}_3}}, \braket{S_{2}^{\text{W}_3}}, \braket{S_{3}^{\text{W}_3}})$  \\ \hline
000  & $(\ket{001}+\ket{010}+\ket{100})/\sqrt{3}$  & $(+1, +1, +1)$      \\ \hline
001  & $(\ket{000}-\ket{011}+\ket{101})/\sqrt{3}$  & $(+1, +1, -1)$   \\ \hline
010  & $(\ket{000}+\ket{011}-\ket{110})/\sqrt{3}$  & $(+1, -1, +1)$     \\ \hline
011  & $(-\ket{001}+\ket{010}-\ket{111})/\sqrt{3}$ & $(+1, -1, -1)$    \\ \hline
100  & $(\ket{000}-\ket{101}+\ket{110})/\sqrt{3}$ & $(-1, +1, +1)$    \\ \hline
101  & $(\ket{001}-\ket{100}-\ket{111})/\sqrt{3}$ & $(-1, +1, -1)$    \\ \hline
110  & $(-\ket{010}+\ket{100}-\ket{111})/\sqrt{3}$ & $(-1, -1, +1)$    \\ \hline
111  & $(-\ket{011}-\ket{101}-\ket{110})/\sqrt{3}$ & $(-1, -1, -1)$    \\ \hline
\end{tabular}
\caption{The $\text{W}_3$-diagonal bases and the corresponding expected values of generators $(\langle S_1^{\text{W}_3}\rangle, \langle S_2^{\text{W}_3}\rangle, \langle S_3^{\text{W}_3}\rangle)$. }
\label{tab:W3bases}
\end{table}
}

Similarly, We can find a possible unitary operator $U^{\text{W}_4}=(ZZZX+ZZX\mathbb{I}+ZX\mathbb{II}+X\mathbb{III})/2$ to generate $\ket{\text{W}_4}$ from $\ket{0000}$. Thus, the generators of $\ket{\text{W}_4}$ can be obtained, i.e.,
\begin{equation}
    \begin{split}
        &S_{1}^{\text{W}_4}=\frac{1}{2}(YZZY +\mathbb{I}YZY +\mathbb{II}YY +\mathbb{III}Z ), \\
        &S_{2}^{\text{W}_4}=\frac{1}{2}(YZY\mathbb{I} +\mathbb{I}YY\mathbb{I} +\mathbb{II}Z\mathbb{I} +\mathbb{II}XX ), \\
        &S_{3}^{\text{W}_4}=\frac{1}{2}(YY\mathbb{II} +\mathbb{I}Z\mathbb{II} +\mathbb{I}XZX +\mathbb{I}XX\mathbb{I} ), \\
        &S_{4}^{\text{W}_4}=\frac{1}{2}(Z\mathbb{III} +XZZX +XZX\mathbb{I} +XX\mathbb{II} ),
    \end{split}
\end{equation}
and other stabilizing operators are
\begin{equation}
    \begin{split}
        &S_5^{\text{W}_4}=\frac{1}{2}(YZYZ +YZ\mathbb{I}Y +\mathbb{I}YYZ +\mathbb{I}Y\mathbb{I}Y),\\
        &S_6^{\text{W}_4}=\frac{1}{2}(YY\mathbb{I}Z +Y\mathbb{I}ZY +\mathbb{I}ZYY +\mathbb{I}XXZ),\\
        &S_7^{\text{W}_4}=\frac{1}{2}(ZYZY +Z\mathbb{I}YY +XZXZ +XX\mathbb{I}Z),\\
        &S_8^{\text{W}_4}=\frac{1}{2}(YYZ\mathbb{I} +Y\mathbb{I}Y\mathbb{I} +\mathbb{I}ZXX+\mathbb{I}X\mathbb{I}X),\\
        &S_9^{\text{W}_4}=\frac{1}{2}(ZYY\mathbb{I} +Z\mathbb{I}XX +XZ\mathbb{I}X +XXZ\mathbb{I}),\\
        &S_{10}^{\text{W}_4}=\frac{1}{2}(ZXZX+ZXX\mathbb{I}+X\mathbb{I}ZX +X\mathbb{I}X\mathbb{I}),\\
        &S_{11}^{\text{W}_4}=\frac{1}{2}(YYZZ +Y\mathbb{I}YZ+Y\mathbb{II}Y-\mathbb{I}ZZZ),\\
        &S_{12}^{\text{W}_4}=\frac{1}{2}(ZYYZ +ZY\mathbb{I}Y -Z\mathbb{I}ZZ +XXZZ),\\
        &S_{13}^{\text{W}_4}=\frac{1}{2}(ZZYY -ZZ\mathbb{I}Z +ZXXZ +X\mathbb{I}XZ),\\
        &S_{14}^{\text{W}_4}=\frac{1}{2}(-ZZZ\mathbb{I} +ZZXX +ZX\mathbb{I}Z +X\mathbb{II}X),\\
        &S_{15}^{\text{W}_4}=-ZZZZ, S_{16}^{\text{W}_4}=\mathbb{IIII}.\\  
    \end{split}
\end{equation}
{The $\text{W}_4$-diagonal bases are shown in \cref{tab:W4bases}.
\begin{table}[h]
	\vspace{0.2cm}
\begin{tabular}{|c|c|c|}
	\hline
	$k_1k_2k_3k_4$     & $\ket{\psi_i^{\text{W}_4}}$     & ($\braket{S_1^{\text{W}_4}}$, $\braket{S_2^{\text{W}_4}}$, $\braket{S_3^{\text{W}_4}}$, $\braket{S_4^{\text{W}_4}}$)  \\ \hline
	0000  & $\frac{\ket{0001}+\ket{0010}+\ket{0100}+\ket{1000}}{2}$ & $(+1, +1, +1 ,+1)$ \\ \hline
	0001  & $\frac{\ket{0000}+\ket{0011}+\ket{0101}+\ket{1001}}{2}$ & $(-1, +1 ,+1, +1)$ \\ \hline
	0010  & $\frac{\ket{0000}-\ket{0011}+\ket{0110}+\ket{1010}}{2}$ & $(+1, -1, +1 ,+1)$ \\ \hline
	0011  & $\frac{\ket{0001}-\ket{0010}+\ket{0111}+\ket{1011}}{2}$ & $(-1, -1, +1 ,+1)$ \\ \hline
	0100  & $\frac{\ket{0000}-\ket{0101}-\ket{0110}+\ket{1100}}{2}$ & $(+1, +1, -1,+1)$ \\ \hline
	0101  & $\frac{\ket{0001}-\ket{0100}-\ket{0111}+\ket{1101}}{2}$ & $(-1, +1, -1,+1)$ \\ \hline
	0110  & $\frac{\ket{0010}-\ket{0100}+\ket{0111}+\ket{1110}}{2}$ & $(1, -1, -1,+1)$ \\ \hline
	0111  & $\frac{\ket{0011}-\ket{0101}+\ket{0110}+\ket{1111}}{2}$ & $(-1, -1, -1,+1)$ \\ \hline
	1000  & $\frac{\ket{0000}-\ket{1001}-\ket{1010}-\ket{1100}}{2}$ & $(+1, +1, +1,-1)$ \\ \hline
	1001  & $\frac{\ket{0001}-\ket{1000}-\ket{1011}-\ket{1101}}{2}$ & $(-1, +1, +1,-1)$ \\ \hline
	1010  & $\frac{\ket{0010}-\ket{1000}+\ket{1011}-\ket{1110}}{2}$ & $(+1, -1, +1,-1)$ \\ \hline
	1011  & $\frac{\ket{0011}-\ket{1001}+\ket{1010}-\ket{1111}}{2}$ & $(-1, -1, +1,-1)$ \\ \hline
	1100  & $\frac{\ket{0100}-\ket{1000}+\ket{1101}+\ket{1110}}{2}$ & $(+1, +1, -1,-1)$ \\ \hline
	1101  & $\frac{\ket{0101}-\ket{1001}+\ket{1100}+\ket{1111}}{2}$ & $(-1, +1, -1,-1)$ \\ \hline
	1110  & $\frac{\ket{0110}-\ket{1010}+\ket{1100}-\ket{1111}}{2}$ & $(+1, -1, -1,-1)$ \\ \hline
	1111  & $\frac{\ket{0111}-\ket{1011}+\ket{1101}-\ket{1110}}{2}$ & $(-1, -1, -1,-1)$ \\ \hline		
\end{tabular}
	\caption{The $\text{W}_4$-diagonal bases and the corresponding expected values of generators $(\langle S_1^{\text{W}_4}\rangle, \langle S_2^{\text{W}_4}\rangle, \langle S_3^{\text{W}_4}\rangle, \langle S_4^{\text{W}_4}\rangle)$. }
	\label{tab:W4bases}
\end{table}
}

\section{Details of experimental realizations and results}
\label{Sec:Expdetails}
In this Appendix, we provide further details about our experimental setup. It would be useful to bear in mind the following:

\begin{enumerate}
\item[(i)] A half-wave plate (HWP) @ $\theta$ performs the unitary transformation $U_\text{\tiny HWP}=\cos2\theta(\proj{H}-\proj{V})+\sin2\theta(\ket{H}\!\bra{V}+\ket{V}\!\bra{H})$ on a polarization state, where $\theta$ is the angle between fast axis of HWP and vertical polarization. \vspace{-0.2cm}
\item[(ii)] A beam displacer (BD) transmits a vertically polarized photon but deviates a horizontally polarized one.
\item[(iii)] A polarized beam splitter (PBS) transmits a horizontally polarized photon but reflects a vertically polarized one.
\item[(iv)] A quarter-wave plate (QWP) @ $\theta$ performs the unitary transformation $U_\text{\tiny QWP}=\frac{1}{\sqrt{2}}[\id_2+{\rm i }\cos2\theta(\proj{H}-\proj{V})+{\rm i }\sin2\theta(\ket{H}\!\bra{V}+\ket{V}\!\bra{H})]$, where $\id_2=\proj{H}+\proj{V}$ and $\theta$ is the angle between fast axis of QWP and vertical polarization, on a polarization state.  
\end{enumerate}

\subsection{Polarization-entangled photon source}
\label{sec:ExpDetail}

\begin{figure}[h!tbp]
\includegraphics[scale=0.25]{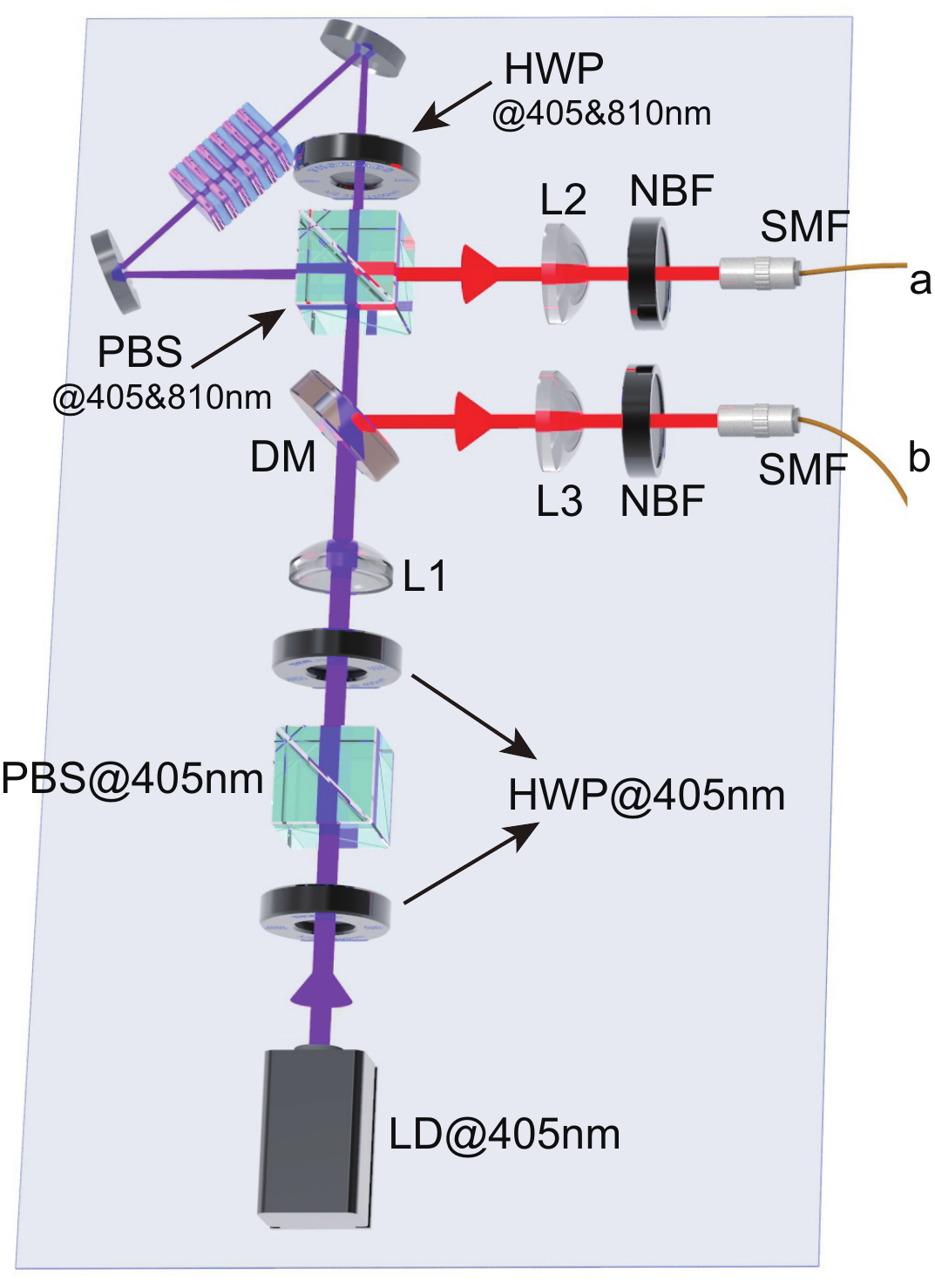} 
\caption{
Illustration of experimental setup to generate polarization-entangled photon pair.}
\label{Fig:PEntS}
\end{figure}
As shown in Fig.~\ref{Fig:PEntS}, the power of the pump light can be adjusted by a HWP and a PBS. After the PBS, horizontal polarization $\ket{H_p}$ is rotated to $\ket{+_p}=\frac{1}{\sqrt{2}}(\ket{H_p}+\ket{V_p})$ by a HWP set at 22.5$^\circ$. Pump beam is focused into PPKTP crystal with beam waist of 74$\mu m$ by two lenses L$_{1}$, whose focal length is 75mm and 125mm respectively. PPKTP crystal, with dimensions of 10 mm (length) $\times$ 2 mm (width) $\times$ 1 mm (thickness) and poling period of $\Lambda=10.025\mu m$, is held in a home-built copper oven and the temperature is controlled by a homemade temperature controller, which is set at 29$^{\circ}$C to realize the optimum type-\uppercase\expandafter{\romannumeral2} phase matching at 810~nm. Then, the pump beam is split by a dual-wavelength PBS, and coherently pumps PPKTP in the clockwise and counterclockwise direction respectively. The clockwise and counterclockwise photons are recombined at the dual-wavelength PBS to generate polarization-entangled photons with ideal form of $\ket{\Psi^{+}_{13}}=\frac{1}{\sqrt{2}}(\ket{H_1V_3}+\ket{V_1H_3})$. Two photons are filtered by narrow band filter (NBF) with full width at half maximum (FWHM) of 3nm, and coupled into single-mode fiber by lenses of focal length 200 mm (L$_2$ and L$_3$) and objective lenses (not shown in Fig.~\ref{Fig:PEntS}). 

\onecolumngrid
\subsection{Experimental setups to generate multi-qubit state}
\label{sec:SetupsMultiQubit}
\begin{figure*}[h!tbp]
\includegraphics[scale=0.7]{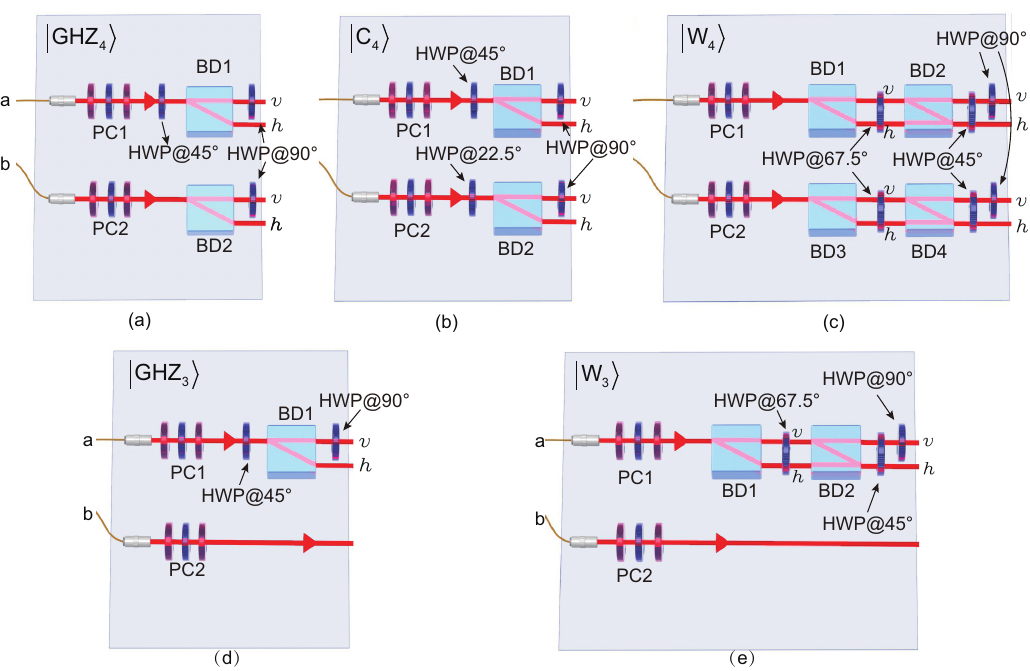} 
\caption{Illustration of the experimental setups to generate multiqubit state (a) the 4-qubit GHZ state $\ket{\text{GHZ}_4}$, (b) the 4-qubit Cluster state $\ket{\text{C}_4}$, (c) the 4-qubit W state $\ket{\text{W}_4}$, (d) the 3-qubit GHZ state $\ket{\text{GHZ}_3}$, and (e) the 3-qubit W state $\ket{\text{W}_4}$.}
\label{Fig:MultiqubitGeneration}
\end{figure*}
We extent photon to its path degree of freedom (DOF) by BD with length of 28.3 mm and clear aperture of 10~mm$\times$10~mm. The BD transmits vertical polarization and deviates horizontal polarization by 3~mm. Specifically, The experimental setups to generate the five multiqubit states $\ket{\text{GHZ}_4}$,  $\ket{\text{C}_4}$, $\ket{\text{W}_4}$, $\ket{\text{GHZ}_3}$ and $\ket{\text{W}_3}$ are shown in Fig.~\ref{Fig:MultiqubitGeneration}(a)-(e), respectively. The step-by-step calculations are shown in \cref{Eq:GHZgeneration}-\cref{eq:Wgeneration}.

\begin{equation}\label{Eq:GHZgeneration}
\begin{split}
\ket{\Psi_{ab}^{+}}&=\frac{1}{\sqrt{2}}(\ket{H_1V_3}+\ket{V_1H_3})\\
&\xrightarrow[\text{on path of photon }a]{\text{HWP}@45^{\circ}}\frac{1}{\sqrt{2}}(\ket{V_1V_3}+\ket{H_1H_3})\\
&\xrightarrow{\text{BD1}}\frac{1}{\sqrt{2}}(\ket{V_1}\ket{v_2}\ket{V_3}+\ket{H_1}\ket{h_2}\ket{H_3})=\ket{\text{GHZ}_3}\\
&\xrightarrow{\text{BD2}}\frac{1}{\sqrt{2}}(\ket{V_1}\ket{v_2}\ket{V_3}\ket{v_4}+\ket{H_1}\ket{h_2}\ket{H_3}\ket{h_4})=\ket{\text{GHZ}_4}\\
\end{split}
\end{equation}

\begin{equation}\label{eq:statePreparation2}
\begin{split}
\ket{\Psi_{ab}^{+}}&=\frac{1}{\sqrt{2}}(\ket{H_1V_3}+\ket{V_1H_3})\\
&\xrightarrow[\text{on path of photon }a]{\text{HWP}@45^{\circ}}\frac{1}{\sqrt{2}}(\ket{V_1V_3}+\ket{H_1H_3})\\
&\xrightarrow[\text{on path of photon }b]{\text{HWP}@22.5^{\circ}}\frac{1}{2}(\ket{V_1H_3}-\ket{V_1V_3}+\ket{H_1H_3}+\ket{H_1V_3})\\
&\xrightarrow{\text{BD1,BD2}}\frac{1}{2}(\ket{V_1}\ket{v_2}\ket{H_3}\ket{h_4}-\ket{V_1}\ket{v_2}\ket{V_3}\ket{v_4}+\ket{H_1}\ket{h_2}\ket{H_3})\ket{h_4}+\ket{H_1}\ket{h_2}\ket{V_3}\ket{v_4})=\ket{\text{C}_4}\\
\end{split}
\end{equation}

	\begin{equation}\label{eq:Wgeneration}
	\begin{aligned}
	\ket{\Psi_{ab}^{+}}&=\frac{1}{\sqrt{2}}(\ket{H_1V_3}+\ket{V_1H_3})\\
	&\xrightarrow{\text{BD1}}\frac{1}{\sqrt{2}}\left(\ket{H_1}\ket{h_2}\ket{V_3}+\ket{V_1}\ket{v_2}\ket{H_3}\right) \\
	&\xrightarrow[\text{on path }h\&v\text{ of photon }a]{\text{HWP}@67.5^{\circ}}\frac{1}{2}\left[(-\ket{H_1}+\ket{V_1})\ket{h_2}\ket{V_3}+(\ket{H_1}+\ket{V_1})\ket{v_2}\ket{H_3} \right] \\
	&\xrightarrow{\text{BD2}}\frac{1}{\sqrt{3}}\left(\ket{V_1}\ket{h_2}\ket{V_3}+\ket{H_1}\ket{h_2}\ket{H_3}+\ket{V_1}\ket{v_2}\ket{H_3} \right) \\
	&\xrightarrow[\text{on path } h\&v\text{ of photon }a]{\text{HWP}@45^{\circ}}\frac{1}{\sqrt{3}}\left( \ket{H_1}\ket{h_2}\ket{V_3}+\ket{V_1}\ket{h_2}\ket{H_3}+\ket{H_1}\ket{v_2}\ket{H_3}\right)=\ket{\text{W}_3}\\
	&\xrightarrow{\text{BD3}}\frac{1}{\sqrt{3}}\left(\ket{H_1}\ket{h_2}\ket{V_3}\ket{v_4}+\ket{V_1}\ket{h_2}\ket{H_3}\ket{h_4}+\ket{H_1}\ket{v_2}\ket{H_3}\ket{h_4}\right) \\
    &\xrightarrow[\text{on path } h\&v \text{ of photon } b]{\text{HWP}@67.5^{\circ}}\frac{1}{\sqrt{6}}\left[\ket{H_1}\ket{h_2}(\ket{H_3}+\ket{V_3})\ket{v_4}+\ket{V_1}\ket{h_2}(-\ket{H_3}+\ket{V_3})\ket{h_4}+\ket{H_1}\ket{v_2}(-\ket{H_3}+\ket{V_3})\ket{h_4}\right]\\
    &\xrightarrow{\text{BD4}}\frac{1}{2}\left(\ket{H_1}\ket{h_2}\ket{H_3}\ket{h_4}+\ket{H_1}\ket{h_2}\ket{V_3}\ket{v_4} +\ket{V_1}\ket{h_2}\ket{V_3}\ket{h_4}+\ket{H_1}\ket{v_2}\ket{V_3}\ket{h_4}\right)\\
	&\xrightarrow[\text{on path }h\&v\text{ of photon }b]{\text{HWP}@45^{\circ}}\frac{1}{2}\left(\ket{H_1}\ket{h_2}\ket{V_3}\ket{h_4}+\ket{H_1}\ket{h_2}\ket{H_3}\ket{v_4}+\ket{V_1}\ket{h_2}\ket{H_3}\ket{h_4}+\ket{H_1}\ket{v_2}\ket{H_3}\ket{h_4}\right)=\ket{\text{W}_4}\\
	\end{aligned}
	\end{equation}

\twocolumngrid
\subsection{Measurement and quantum state tomography}

\begin{figure}[htbp]
\includegraphics[width=0.8\columnwidth]{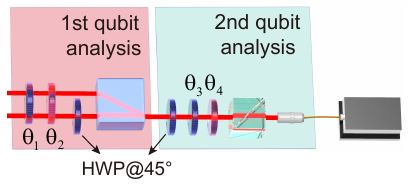} 
\caption{Experimental setups of measurement on a photon encoded in polarization DOF and path DOF.}
\label{Fig:MeasureSetup}
\end{figure}

If the photon is encoded either in polarization DOF or path DOF, we first perform measurement on polarization DOF and then the path DOF. As illustrated in \cref{Fig:MeasureSetup}, the measurement basis of qubit on polarization DOF $\alpha\ket{H}+\beta\ket{V}$ is determined by HWP at $\theta_1$ and QWP at $\theta_2$. The measurement basis of qubit on path DOF $\gamma\ket{h}+\delta\ket{v}$ is determined by HWP at $\theta_3$ and QWP at $\theta_4$. Finally, a PBS is applied before the photon arrives detector. With this setting, the measurement on basis $(\alpha\ket{H}+\beta\ket{V})\otimes(\gamma\ket{h}+\delta\ket{v})$ is achieved. The specific calculations are shown in \cref{Eq:Measuresetup}. If the photon is only encoded in polarization DOF, the measurement on basis $\alpha\ket{H}+\beta\ket{V}$ is implemented by a QWP, HWP and PBS. 

\begin{equation}\label{Eq:Measuresetup}
\begin{split}
&\left(\alpha\ket{H}+\beta\ket{V}\right)\otimes\left(\gamma\ket{h}+\delta\ket{v}\right)\\
&\xrightarrow[\text{QWP}@{\theta}_2]{\text{HWP}@{\theta}_1}\ket{H}\otimes(\gamma\ket{h}+\delta\ket{v})\\
&\xrightarrow[\text{on path }h]{\text{HWP}@45^{\circ}}\gamma\ket{V}\ket{h}+\delta\ket{H}\ket{v} \\
&\xrightarrow{\text{BD}}\gamma\ket{V}\ket{h}+\delta\ket{H}\ket{h}\\
&\xrightarrow{\text{HWP}@45^{\circ}}(\gamma\ket{H}+\delta\ket{V})\ket{h} \\
&\xrightarrow[\text{QWP}@{\theta}_4]{\text{HWP}@{\theta}_3}\ket{H}\ket{h}\\
&\xrightarrow{\text{PBS}}\ket{H}\ket{h}\\
\end{split}
\end{equation}

\begin{figure*}[htbp]
\includegraphics[width=1.8\columnwidth]{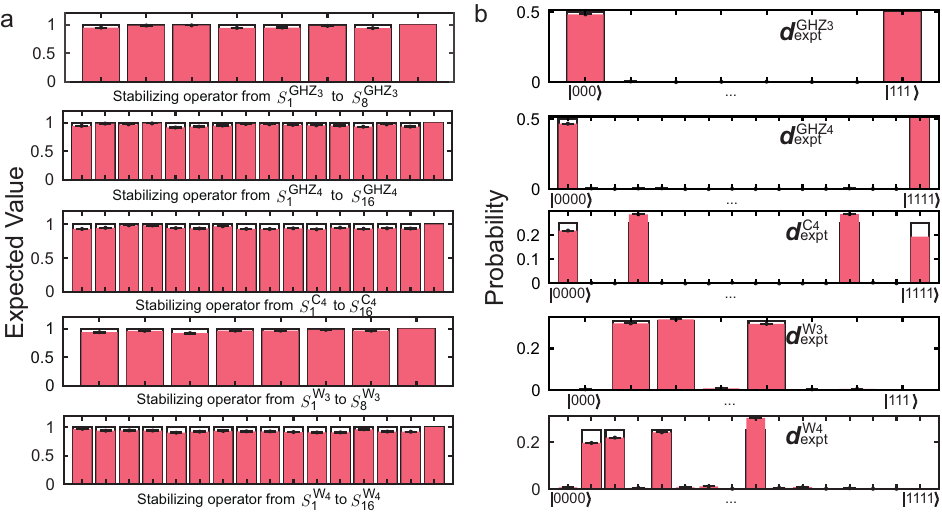}
\caption{Experimental results of (a) expected values of $\langle S_i^{\psi}\rangle$ and (b) probabilities of $\bm{d}_{\text{expt}}^{\psi}$ for state $\rho_{\text{expt}}^{\text{GHZ}_3}$, $\rho_{\text{expt}}^{\text{GHZ}_4}$, $\rho_{\text{expt}}^{\text{C}_4}$, $\rho_{\text{expt}}^{\text{W}_3}$ and $\rho_{\text{expt}}^{\text{W}_4}$, respectively. The black grids represent the values for ideal states.}
\label{Fig:ExpectandProbZ}
\end{figure*}

\begin{figure*}[htbp]
\includegraphics[width=1\columnwidth]{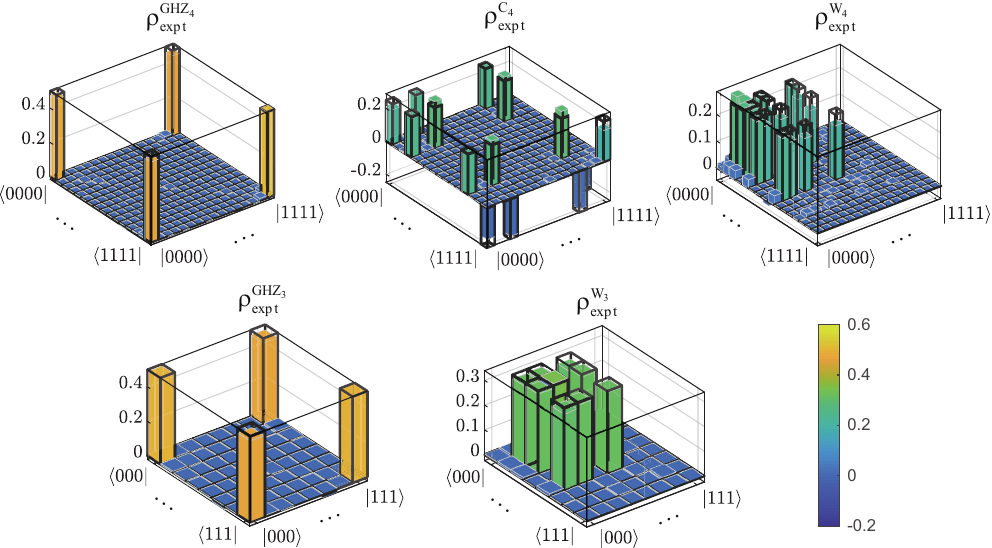} 
\caption{Reconstructed density matrices of generated multi-qubit states. }
\label{Fig:StateTomo}
\end{figure*}

With this experimental setting, we can perform measurement on arbitrary basis. The experimental results of $\langle S_i^{\psi}\rangle$ and  $\bm{d}_{\text{expt}}^{\psi}$ are shown in Fig.~\ref{Fig:ExpectandProbZ}a and b, respectively. Moreover, we reconstruct experimentally generated states $\rho_{\text{expt}}^{\text{GHZ}_4}$, $\rho_{\text{expt}}^{\text{C}_4}$, $\rho_{\text{expt}}^{\text{W}_4}$, $\rho_{\text{expt}}^{\text{GHZ}_3}$ and $\rho_{\text{expt}}^{\text{W}_3}$ by quantum state tomography\cite{nielsen_quantum_2010}. The results are shown in \cref{Fig:StateTomo}, from which we calculate the fidelity $F^{\psi}=\tr(\rho_{\text{expt}}^{\psi}\ket{\psi}\bra{\psi})$. We observe that $F^{\text{GHZ}_3}=0.9643\pm0.0003$, $F^{\text{W}_3}=0.9589\pm0.0005$, $F^{\text{GHZ}_4}=0.9571\pm0.0003$, $F^{\text{C}_4}=0.9497\pm0.0002$ and $F^{\text{W}_4}=0.915\pm0.001$. Also, the relative entropy of coherence $C_{\text{RE}}(\rho_{\text{expt}}^{\psi})$ of $\rho_{\text{expt}}^{\psi}$ can be calculated by $C_{\text{RE}}(\rho)=S_{\text{VN}}(\rho_{d})-S_{\text{VN}}(\rho)$, according to which we obtain $C_{\text{RE}}(\rho_{\text{expt}}^{\text{GHZ}_3})=0.875\pm0.002$,  $C_{\text{RE}}(\rho_{\text{expt}}^{\text{W}_3})=1.479\pm0.004$,  $C_{\text{RE}}(\rho_{\text{expt}}^{\text{GHZ}_4})=0.906\pm0.002$,  $C_{\text{RE}}(\rho_{\text{expt}}^{\text{C}_4})=1.806\pm0.002$ and  $C_{\text{RE}}(\rho_{\text{expt}}^{\text{W}_4})=1.964\pm0.003$.

\onecolumngrid

{
\section{Experimental imperfections}
\label{Sec:Noise}
In our experiment, the main imperfection of prepared multi-qubit states comes from the imperfection of the polarization entangled-photon pair, which is caused by the mode mismatch of overlapping lights on PBS in Fig.~\ref{Fig:PEntS}. The noisy state can be described by a dephasing channel acting on ideal $\ket{\Phi^+}=(\ket{HH}+\ket{VV})/\sqrt{2}$ by   
\begin{equation}
\rho^{\Psi^+}=\mathcal E (\ket{\Phi^+}\bra{\Phi^+})=(1-\mu)\ket{\Phi^+}\bra{\Phi^+}+\frac{\mu}{2}(\ket{HH}\bra{HH}+\ket{VV}\bra{VV}).
\end{equation}
With such a noisy state and following the procedure of generating multi-qubit state in Appendix~\ref{sec:SetupsMultiQubit}, we calculate the noisy multi-qubit states in computational basis (CB)
\begin{align}
&\rho^{\text{GHZ}_3}_{\text{CB}}=(1-\mu)\ket{\text{GHZ}_3}\bra{\text{GHZ}_3}+\frac{\mu}{2}(\ket{000}\bra{000}+\ket{111}\bra{111}),\\
&\rho^{\text{GHZ}_4}_{\text{CB}}=(1-\mu)\ket{\text{GHZ}_4}\bra{\text{GHZ}_4}+\frac{\mu}{2}(\ket{0000}\bra{0000}+\ket{1111}\bra{1111}),\\
&\rho^{\text{C}_4}_{\text{CB}}=(1-\mu)\ket{\text{C}_4}\bra{\text{C}_4}+\frac{\mu}{4}(\ket{0000}\bra{0000}+\ket{0011}\bra{0011}+\ket{1100}\bra{1100}+\ket{1111}\bra{1111}),\\
&\rho^{\text{W}_3}_{\text{CB}}=(1-\mu)\ket{\text{W}_3}\bra{\text{W}_3}+\frac{\mu}{3}(\ket{100}\bra{100}+\ket{010}\bra{010}+\ket{001}\bra{001}),\\
&\rho^{\text{W}_4}_{\text{CB}}=(1-\mu)\ket{\text{W}_4}\bra{\text{W}_4}+\frac{\mu}{4}(\ket{1000}\bra{1000}+\ket{0100}\bra{0100}+\ket{0010}\bra{0010}+\ket{0001}\bra{0001}).
\end{align}
In the graph-diagonal basis (GDB) ,
\begin{align}
&\rho^{\text{GHZ}_3}_{\text{GDB}}=U^{\text{GHZ}_3}\rho^{\text{GHZ}_3}_{\text{CB}}{U^{\text{GHZ}_3}}^\dagger=\sum_{k=1}^8\lambda_k\ket{\psi^{\text{GHZ}_3}_k}\bra{\psi^{\text{GHZ}_3}_k}=(1-\frac{\mu}{2})\ket{\psi^{\text{GHZ}_3}_1}\bra{\psi^{\text{GHZ}_3}_1}+\frac{\mu}{2}\ket{\psi^{\text{GHZ}_3}_{8}}\bra{\psi^{\text{GHZ}_3}_{8}},\\
&\rho^{\text{GHZ}_4}_{\text{GDB}}=U^{\text{GHZ}_4}\rho^{\text{GHZ}_4}_{\text{CB}}{U^{\text{GHZ}_4}}^\dagger=\sum_{k=1}^{16}\lambda_k\ket{\psi^{\text{GHZ}_4}_k}\bra{\psi^{\text{GHZ}_4}_k}=(1-\frac{\mu}{2})\ket{\psi^{\text{GHZ}_4}_1}\bra{\psi^{\text{GHZ}_4}_1}+\frac{\mu}{2}\ket{\psi^{\text{GHZ}_4}_{16}}\bra{\psi^{\text{GHZ}_4}_{16}},\\
&\rho^{\text{C}_4}_{\text{GDB}}=U^{\text{C}_4}\rho^{\text{C}_4}_{\text{CB}}{U^{\text{C}_4}}^\dagger=\sum_{k=1}^{16}\lambda_k\ket{\psi^{\text{C}_4}_k}\bra{\psi^{\text{C}_4}_k}=(1-\frac{3\mu}{4})\ket{\psi^{\text{C}_4}_1}\bra{\psi^{\text{C}_4}_1}+\frac{\mu}{4}(\ket{\psi^{\text{C}_4}_{4}}\bra{\psi^{\text{C}_4}_{4}}+\ket{\psi^{\text{C}_4}_{13}}\bra{\psi^{\text{C}_4}_{13}}+\ket{\psi^{\text{C}_4}_{16}}\bra{\psi^{\text{C}_4}_{16}})
\end{align}
are still graph-diagonal states under dephasing channel, which means that the estimated lower bounds on such states are tight, and it is easy to check that $\rho^{\text{W}_3}_{\text{GDB}}=U^{\text{W}_3}\rho^{\text{W}_3}_{\text{CB}}{U^{\text{W}_3}}^\dagger\neq\sum_k\lambda_k\ket{\psi^{\text{W}_3}_k}\bra{\psi^{\text{W}_3}_k}$ and $\rho^{\text{W}_4}_{\text{GDB}}=U^{\text{W}_4}\rho^{\text{W}_4}_{\text{CB}}{U^{\text{W}_4}}^\dagger\neq\sum_k\lambda_k\ket{\psi^{\text{W}_4}_k}\bra{\psi^{\text{W}_4}_k}$.

In experiment, we observe that the normalized distance of estimated $l^c(\rho^{\text{W}_4})$ is larger than that of other states as the prepared $\rho_{\text{expt}}^{\text{W}_4}$ has lower fidelity compared to other states. More importantly, it is not a graph-diagonal state any more under dephasing channel, which causes that there is only 5\% difference in state fidelity (shown in \cref{Sec:Expdetails}) but 15\% difference in normalized distance (shown in \cref{Fig:data2}c).}
\twocolumngrid
\bibliography{MCE_Arxiv.bbl}

\begin{thebibliography}{67}%
\makeatletter
\providecommand \@ifxundefined [1]{%
 \@ifx{#1\undefined}
}%
\providecommand \@ifnum [1]{%
 \ifnum #1\expandafter \@firstoftwo
 \else \expandafter \@secondoftwo
 \fi
}%
\providecommand \@ifx [1]{%
 \ifx #1\expandafter \@firstoftwo
 \else \expandafter \@secondoftwo
 \fi
}%
\providecommand \natexlab [1]{#1}%
\providecommand \enquote  [1]{``#1''}%
\providecommand \bibnamefont  [1]{#1}%
\providecommand \bibfnamefont [1]{#1}%
\providecommand \citenamefont [1]{#1}%
\providecommand \href@noop [0]{\@secondoftwo}%
\providecommand \href [0]{\begingroup \@sanitize@url \@href}%
\providecommand \@href[1]{\@@startlink{#1}\@@href}%
\providecommand \@@href[1]{\endgroup#1\@@endlink}%
\providecommand \@sanitize@url [0]{\catcode `\\12\catcode `\$12\catcode
  `\&12\catcode `\#12\catcode `\^12\catcode `\_12\catcode `\%12\relax}%
\providecommand \@@startlink[1]{}%
\providecommand \@@endlink[0]{}%
\providecommand \url  [0]{\begingroup\@sanitize@url \@url }%
\providecommand \@url [1]{\endgroup\@href {#1}{\urlprefix }}%
\providecommand \urlprefix  [0]{URL }%
\providecommand \Eprint [0]{\href }%
\providecommand \doibase [0]{http://dx.doi.org/}%
\providecommand \selectlanguage [0]{\@gobble}%
\providecommand \bibinfo  [0]{\@secondoftwo}%
\providecommand \bibfield  [0]{\@secondoftwo}%
\providecommand \translation [1]{[#1]}%
\providecommand \BibitemOpen [0]{}%
\providecommand \bibitemStop [0]{}%
\providecommand \bibitemNoStop [0]{.\EOS\space}%
\providecommand \EOS [0]{\spacefactor3000\relax}%
\providecommand \BibitemShut  [1]{\csname bibitem#1\endcsname}%
\let\auto@bib@innerbib\@empty
\bibitem [{\citenamefont {Grosshans}\ \emph {et~al.}(2003)\citenamefont
  {Grosshans}, \citenamefont {Van~Assche}, \citenamefont {Wenger},
  \citenamefont {Brouri}, \citenamefont {Cerf},\ and\ \citenamefont
  {Grangier}}]{Grosshans03}%
  \BibitemOpen
  \bibfield  {author} {\bibinfo {author} {\bibfnamefont {F.}~\bibnamefont
  {Grosshans}}, \bibinfo {author} {\bibfnamefont {G.}~\bibnamefont
  {Van~Assche}}, \bibinfo {author} {\bibfnamefont {J.}~\bibnamefont {Wenger}},
  \bibinfo {author} {\bibfnamefont {R.}~\bibnamefont {Brouri}}, \bibinfo
  {author} {\bibfnamefont {N.~J.}\ \bibnamefont {Cerf}}, \ and\ \bibinfo
  {author} {\bibfnamefont {P.}~\bibnamefont {Grangier}},\ }\href@noop {}
  {\bibfield  {journal} {\bibinfo  {journal} {Nature}\ }\textbf {\bibinfo
  {volume} {421}},\ \bibinfo {pages} {238} (\bibinfo {year}
  {2003})}\BibitemShut {NoStop}%
\bibitem [{\citenamefont {Coles}\ \emph {et~al.}(2016)\citenamefont {Coles},
  \citenamefont {Metodiev},\ and\ \citenamefont
  {L{\"u}tkenhaus}}]{coles2016numerical}%
  \BibitemOpen
  \bibfield  {author} {\bibinfo {author} {\bibfnamefont {P.~J.}\ \bibnamefont
  {Coles}}, \bibinfo {author} {\bibfnamefont {E.~M.}\ \bibnamefont {Metodiev}},
  \ and\ \bibinfo {author} {\bibfnamefont {N.}~\bibnamefont {L{\"u}tkenhaus}},\
  }\href@noop {} {\bibfield  {journal} {\bibinfo  {journal} {Nature
  communications}\ }\textbf {\bibinfo {volume} {7}},\ \bibinfo {pages} {1}
  (\bibinfo {year} {2016})}\BibitemShut {NoStop}%
\bibitem [{\citenamefont {Ma}\ \emph {et~al.}(2019)\citenamefont {Ma},
  \citenamefont {Zhou}, \citenamefont {Yuan},\ and\ \citenamefont
  {Ma}}]{ma2019operational}%
  \BibitemOpen
  \bibfield  {author} {\bibinfo {author} {\bibfnamefont {J.}~\bibnamefont
  {Ma}}, \bibinfo {author} {\bibfnamefont {Y.}~\bibnamefont {Zhou}}, \bibinfo
  {author} {\bibfnamefont {X.}~\bibnamefont {Yuan}}, \ and\ \bibinfo {author}
  {\bibfnamefont {X.}~\bibnamefont {Ma}},\ }\href {\doibase
  10.1103/PhysRevA.99.062325} {\bibfield  {journal} {\bibinfo  {journal} {Phys.
  Rev. A}\ }\textbf {\bibinfo {volume} {99}},\ \bibinfo {pages} {062325}
  (\bibinfo {year} {2019})}\BibitemShut {NoStop}%
\bibitem [{\citenamefont {Giovannetti}\ \emph {et~al.}(2011)\citenamefont
  {Giovannetti}, \citenamefont {Lloyd},\ and\ \citenamefont
  {Maccone}}]{Giovannetti2011}%
  \BibitemOpen
  \bibfield  {author} {\bibinfo {author} {\bibfnamefont {V.}~\bibnamefont
  {Giovannetti}}, \bibinfo {author} {\bibfnamefont {S.}~\bibnamefont {Lloyd}},
  \ and\ \bibinfo {author} {\bibfnamefont {L.}~\bibnamefont {Maccone}},\
  }\href@noop {} {\bibfield  {journal} {\bibinfo  {journal} {Nature Photonics}\
  }\textbf {\bibinfo {volume} {5}},\ \bibinfo {pages} {222} (\bibinfo {year}
  {2011})}\BibitemShut {NoStop}%
\bibitem [{\citenamefont {Zhang}\ \emph {et~al.}(2019)\citenamefont {Zhang},
  \citenamefont {Bromley}, \citenamefont {Huang}, \citenamefont {Cao},
  \citenamefont {Lv}, \citenamefont {Liu}, \citenamefont {Li}, \citenamefont
  {Guo}, \citenamefont {Cianciaruso},\ and\ \citenamefont
  {Adesso}}]{zhang2019demonstrating}%
  \BibitemOpen
  \bibfield  {author} {\bibinfo {author} {\bibfnamefont {C.}~\bibnamefont
  {Zhang}}, \bibinfo {author} {\bibfnamefont {T.~R.}\ \bibnamefont {Bromley}},
  \bibinfo {author} {\bibfnamefont {Y.-F.}\ \bibnamefont {Huang}}, \bibinfo
  {author} {\bibfnamefont {H.}~\bibnamefont {Cao}}, \bibinfo {author}
  {\bibfnamefont {W.-M.}\ \bibnamefont {Lv}}, \bibinfo {author} {\bibfnamefont
  {B.-H.}\ \bibnamefont {Liu}}, \bibinfo {author} {\bibfnamefont {C.-F.}\
  \bibnamefont {Li}}, \bibinfo {author} {\bibfnamefont {G.-C.}\ \bibnamefont
  {Guo}}, \bibinfo {author} {\bibfnamefont {M.}~\bibnamefont {Cianciaruso}}, \
  and\ \bibinfo {author} {\bibfnamefont {G.}~\bibnamefont {Adesso}},\ }\href
  {\doibase 10.1103/PhysRevLett.123.180504} {\bibfield  {journal} {\bibinfo
  {journal} {Phys. Rev. Lett.}\ }\textbf {\bibinfo {volume} {123}},\ \bibinfo
  {pages} {180504} (\bibinfo {year} {2019})}\BibitemShut {NoStop}%
\bibitem [{\citenamefont {\AA{}berg}(2014)}]{aaberg2014catalytic}%
  \BibitemOpen
  \bibfield  {author} {\bibinfo {author} {\bibfnamefont {J.}~\bibnamefont
  {\AA{}berg}},\ }\href {\doibase 10.1103/PhysRevLett.113.150402} {\bibfield
  {journal} {\bibinfo  {journal} {Phys. Rev. Lett.}\ }\textbf {\bibinfo
  {volume} {113}},\ \bibinfo {pages} {150402} (\bibinfo {year}
  {2014})}\BibitemShut {NoStop}%
\bibitem [{\citenamefont {Lostaglio}\ \emph {et~al.}(2015)\citenamefont
  {Lostaglio}, \citenamefont {Jennings},\ and\ \citenamefont
  {Rudolph}}]{Lostaglio2015}%
  \BibitemOpen
  \bibfield  {author} {\bibinfo {author} {\bibfnamefont {M.}~\bibnamefont
  {Lostaglio}}, \bibinfo {author} {\bibfnamefont {D.}~\bibnamefont {Jennings}},
  \ and\ \bibinfo {author} {\bibfnamefont {T.}~\bibnamefont {Rudolph}},\
  }\href@noop {} {\bibfield  {journal} {\bibinfo  {journal} {Nature
  Communications}\ }\textbf {\bibinfo {volume} {6}},\ \bibinfo {pages} {6383}
  (\bibinfo {year} {2015})}\BibitemShut {NoStop}%
\bibitem [{\citenamefont {Narasimhachar}\ and\ \citenamefont
  {Gour}(2015)}]{Narasimhachar2015}%
  \BibitemOpen
  \bibfield  {author} {\bibinfo {author} {\bibfnamefont {V.}~\bibnamefont
  {Narasimhachar}}\ and\ \bibinfo {author} {\bibfnamefont {G.}~\bibnamefont
  {Gour}},\ }\href@noop {} {\bibfield  {journal} {\bibinfo  {journal} {Nature
  Communications}\ }\textbf {\bibinfo {volume} {6}},\ \bibinfo {pages} {7689}
  (\bibinfo {year} {2015})}\BibitemShut {NoStop}%
\bibitem [{\citenamefont {Romero}\ \emph {et~al.}(2014)\citenamefont {Romero},
  \citenamefont {Augulis}, \citenamefont {Novoderezhkin}, \citenamefont
  {Ferretti}, \citenamefont {Thieme}, \citenamefont {Zigmantas},\ and\
  \citenamefont {van Grondelle}}]{Romero2014}%
  \BibitemOpen
  \bibfield  {author} {\bibinfo {author} {\bibfnamefont {E.}~\bibnamefont
  {Romero}}, \bibinfo {author} {\bibfnamefont {R.}~\bibnamefont {Augulis}},
  \bibinfo {author} {\bibfnamefont {V.~I.}\ \bibnamefont {Novoderezhkin}},
  \bibinfo {author} {\bibfnamefont {M.}~\bibnamefont {Ferretti}}, \bibinfo
  {author} {\bibfnamefont {J.}~\bibnamefont {Thieme}}, \bibinfo {author}
  {\bibfnamefont {D.}~\bibnamefont {Zigmantas}}, \ and\ \bibinfo {author}
  {\bibfnamefont {R.}~\bibnamefont {van Grondelle}},\ }\href@noop {} {\bibfield
   {journal} {\bibinfo  {journal} {Nature Physics}\ }\textbf {\bibinfo {volume}
  {10}},\ \bibinfo {pages} {676} (\bibinfo {year} {2014})}\BibitemShut
  {NoStop}%
\bibitem [{\citenamefont {Vedral}\ \emph {et~al.}(1997)\citenamefont {Vedral},
  \citenamefont {Plenio}, \citenamefont {Rippin},\ and\ \citenamefont
  {Knight}}]{Vedral1997}%
  \BibitemOpen
  \bibfield  {author} {\bibinfo {author} {\bibfnamefont {V.}~\bibnamefont
  {Vedral}}, \bibinfo {author} {\bibfnamefont {M.~B.}\ \bibnamefont {Plenio}},
  \bibinfo {author} {\bibfnamefont {M.~A.}\ \bibnamefont {Rippin}}, \ and\
  \bibinfo {author} {\bibfnamefont {P.~L.}\ \bibnamefont {Knight}},\ }\href
  {\doibase 10.1103/PhysRevLett.78.2275} {\bibfield  {journal} {\bibinfo
  {journal} {Phys. Rev. Lett.}\ }\textbf {\bibinfo {volume} {78}},\ \bibinfo
  {pages} {2275} (\bibinfo {year} {1997})}\BibitemShut {NoStop}%
\bibitem [{\citenamefont {Vedral}\ and\ \citenamefont
  {Plenio}(1998)}]{Vedral1998}%
  \BibitemOpen
  \bibfield  {author} {\bibinfo {author} {\bibfnamefont {V.}~\bibnamefont
  {Vedral}}\ and\ \bibinfo {author} {\bibfnamefont {M.~B.}\ \bibnamefont
  {Plenio}},\ }\href {\doibase 10.1103/PhysRevA.57.1619} {\bibfield  {journal}
  {\bibinfo  {journal} {Phys. Rev. A}\ }\textbf {\bibinfo {volume} {57}},\
  \bibinfo {pages} {1619} (\bibinfo {year} {1998})}\BibitemShut {NoStop}%
\bibitem [{\citenamefont {Chitambar}\ and\ \citenamefont
  {Gour}(2019)}]{RevModPhys.91.025001}%
  \BibitemOpen
  \bibfield  {author} {\bibinfo {author} {\bibfnamefont {E.}~\bibnamefont
  {Chitambar}}\ and\ \bibinfo {author} {\bibfnamefont {G.}~\bibnamefont
  {Gour}},\ }\href {\doibase 10.1103/RevModPhys.91.025001} {\bibfield
  {journal} {\bibinfo  {journal} {Rev. Mod. Phys.}\ }\textbf {\bibinfo {volume}
  {91}},\ \bibinfo {pages} {025001} (\bibinfo {year} {2019})}\BibitemShut
  {NoStop}%
\bibitem [{\citenamefont {Aberg}(2006)}]{aberg2006}%
  \BibitemOpen
  \bibfield  {author} {\bibinfo {author} {\bibfnamefont {J.}~\bibnamefont
  {Aberg}},\ }\href@noop {} {\enquote {\bibinfo {title} {Quantifying
  superposition},}\ } (\bibinfo {year} {2006}),\ \Eprint
  {http://arxiv.org/abs/quant-ph/0612146} {arXiv:quant-ph/0612146 [quant-ph]}
  \BibitemShut {NoStop}%
\bibitem [{\citenamefont {Baumgratz}\ \emph {et~al.}(2014)\citenamefont
  {Baumgratz}, \citenamefont {Cramer},\ and\ \citenamefont
  {Plenio}}]{Baumgratz2014}%
  \BibitemOpen
  \bibfield  {author} {\bibinfo {author} {\bibfnamefont {T.}~\bibnamefont
  {Baumgratz}}, \bibinfo {author} {\bibfnamefont {M.}~\bibnamefont {Cramer}}, \
  and\ \bibinfo {author} {\bibfnamefont {M.~B.}\ \bibnamefont {Plenio}},\
  }\href {\doibase 10.1103/PhysRevLett.113.140401} {\bibfield  {journal}
  {\bibinfo  {journal} {Phys. Rev. Lett.}\ }\textbf {\bibinfo {volume} {113}},\
  \bibinfo {pages} {140401} (\bibinfo {year} {2014})}\BibitemShut {NoStop}%
\bibitem [{\citenamefont {Streltsov}\ \emph {et~al.}(2015)\citenamefont
  {Streltsov}, \citenamefont {Singh}, \citenamefont {Dhar}, \citenamefont
  {Bera},\ and\ \citenamefont {Adesso}}]{streltsov_measuring_2015}%
  \BibitemOpen
  \bibfield  {author} {\bibinfo {author} {\bibfnamefont {A.}~\bibnamefont
  {Streltsov}}, \bibinfo {author} {\bibfnamefont {U.}~\bibnamefont {Singh}},
  \bibinfo {author} {\bibfnamefont {H.~S.}\ \bibnamefont {Dhar}}, \bibinfo
  {author} {\bibfnamefont {M.~N.}\ \bibnamefont {Bera}}, \ and\ \bibinfo
  {author} {\bibfnamefont {G.}~\bibnamefont {Adesso}},\ }\href {\doibase
  10.1103/PhysRevLett.115.020403} {\bibfield  {journal} {\bibinfo  {journal}
  {Phys. Rev. Lett.}\ }\textbf {\bibinfo {volume} {115}},\ \bibinfo {pages}
  {020403} (\bibinfo {year} {2015})}\BibitemShut {NoStop}%
\bibitem [{\citenamefont {Napoli}\ \emph {et~al.}(2016)\citenamefont {Napoli},
  \citenamefont {Bromley}, \citenamefont {Cianciaruso}, \citenamefont {Piani},
  \citenamefont {Johnston},\ and\ \citenamefont
  {Adesso}}]{napoli_robustness_2016}%
  \BibitemOpen
  \bibfield  {author} {\bibinfo {author} {\bibfnamefont {C.}~\bibnamefont
  {Napoli}}, \bibinfo {author} {\bibfnamefont {T.~R.}\ \bibnamefont {Bromley}},
  \bibinfo {author} {\bibfnamefont {M.}~\bibnamefont {Cianciaruso}}, \bibinfo
  {author} {\bibfnamefont {M.}~\bibnamefont {Piani}}, \bibinfo {author}
  {\bibfnamefont {N.}~\bibnamefont {Johnston}}, \ and\ \bibinfo {author}
  {\bibfnamefont {G.}~\bibnamefont {Adesso}},\ }\href {\doibase
  10.1103/PhysRevLett.116.150502} {\bibfield  {journal} {\bibinfo  {journal}
  {Phys. Rev. Lett.}\ }\textbf {\bibinfo {volume} {116}},\ \bibinfo {pages}
  {150502} (\bibinfo {year} {2016})}\BibitemShut {NoStop}%
\bibitem [{\citenamefont {Bu}\ \emph {et~al.}(2017)\citenamefont {Bu},
  \citenamefont {Singh}, \citenamefont {Fei}, \citenamefont {Pati},\ and\
  \citenamefont {Wu}}]{bu_maximum_2017}%
  \BibitemOpen
  \bibfield  {author} {\bibinfo {author} {\bibfnamefont {K.}~\bibnamefont
  {Bu}}, \bibinfo {author} {\bibfnamefont {U.}~\bibnamefont {Singh}}, \bibinfo
  {author} {\bibfnamefont {S.-M.}\ \bibnamefont {Fei}}, \bibinfo {author}
  {\bibfnamefont {A.~K.}\ \bibnamefont {Pati}}, \ and\ \bibinfo {author}
  {\bibfnamefont {J.}~\bibnamefont {Wu}},\ }\href {\doibase
  10.1103/PhysRevLett.119.150405} {\bibfield  {journal} {\bibinfo  {journal}
  {Phys. Rev. Lett.}\ }\textbf {\bibinfo {volume} {119}},\ \bibinfo {pages}
  {150405} (\bibinfo {year} {2017})}\BibitemShut {NoStop}%
\bibitem [{\citenamefont {Chen}\ and\ \citenamefont
  {Fei}(2018)}]{chen_notes_2018}%
  \BibitemOpen
  \bibfield  {author} {\bibinfo {author} {\bibfnamefont {B.}~\bibnamefont
  {Chen}}\ and\ \bibinfo {author} {\bibfnamefont {S.-M.}\ \bibnamefont {Fei}},\
  }\href {\doibase 10.1007/s11128-018-1879-9} {\bibfield  {journal} {\bibinfo
  {journal} {Quantum Inf Process}\ }\textbf {\bibinfo {volume} {17}},\ \bibinfo
  {pages} {107} (\bibinfo {year} {2018})}\BibitemShut {NoStop}%
\bibitem [{\citenamefont {Du}\ \emph {et~al.}(2019)\citenamefont {Du},
  \citenamefont {Bai},\ and\ \citenamefont {Qi}}]{du_coherence_2019}%
  \BibitemOpen
  \bibfield  {author} {\bibinfo {author} {\bibfnamefont {S.}~\bibnamefont
  {Du}}, \bibinfo {author} {\bibfnamefont {Z.}~\bibnamefont {Bai}}, \ and\
  \bibinfo {author} {\bibfnamefont {X.}~\bibnamefont {Qi}},\ }\href {\doibase
  10.1103/PhysRevA.100.032313} {\bibfield  {journal} {\bibinfo  {journal}
  {Phys. Rev. A}\ }\textbf {\bibinfo {volume} {100}},\ \bibinfo {pages}
  {032313} (\bibinfo {year} {2019})}\BibitemShut {NoStop}%
\bibitem [{\citenamefont {Du}\ \emph {et~al.}(2015)\citenamefont {Du},
  \citenamefont {Bai},\ and\ \citenamefont {Qi}}]{du_coherence_2015}%
  \BibitemOpen
  \bibfield  {author} {\bibinfo {author} {\bibfnamefont {S.}~\bibnamefont
  {Du}}, \bibinfo {author} {\bibfnamefont {Z.}~\bibnamefont {Bai}}, \ and\
  \bibinfo {author} {\bibfnamefont {X.}~\bibnamefont {Qi}},\ }\href@noop {}
  {\bibfield  {journal} {\bibinfo  {journal} {Quantum Inf.Comput.}\ }\bibinfo
  {series} {15},\ \textbf {\bibinfo {volume} {15}},\ \bibinfo {pages} {1307}
  (\bibinfo {year} {2015})}\BibitemShut {NoStop}%
\bibitem [{\citenamefont {Liu}\ \emph {et~al.}(2017{\natexlab{a}})\citenamefont
  {Liu}, \citenamefont {Zhang}, \citenamefont {Yu}, \citenamefont {Ding},\ and\
  \citenamefont {Liu}}]{liu_new_2017}%
  \BibitemOpen
  \bibfield  {author} {\bibinfo {author} {\bibfnamefont {C.~L.}\ \bibnamefont
  {Liu}}, \bibinfo {author} {\bibfnamefont {D.-J.}\ \bibnamefont {Zhang}},
  \bibinfo {author} {\bibfnamefont {X.-D.}\ \bibnamefont {Yu}}, \bibinfo
  {author} {\bibfnamefont {Q.-M.}\ \bibnamefont {Ding}}, \ and\ \bibinfo
  {author} {\bibfnamefont {L.}~\bibnamefont {Liu}},\ }\href {\doibase
  10.1007/s11128-017-1650-7} {\bibfield  {journal} {\bibinfo  {journal}
  {Quantum Inf Process}\ }\textbf {\bibinfo {volume} {16}},\ \bibinfo {pages}
  {198} (\bibinfo {year} {2017}{\natexlab{a}})}\BibitemShut {NoStop}%
\bibitem [{\citenamefont {Qi}\ \emph {et~al.}(2017)\citenamefont {Qi},
  \citenamefont {Gao},\ and\ \citenamefont {Yan}}]{qi_measuring_2017}%
  \BibitemOpen
  \bibfield  {author} {\bibinfo {author} {\bibfnamefont {X.}~\bibnamefont
  {Qi}}, \bibinfo {author} {\bibfnamefont {T.}~\bibnamefont {Gao}}, \ and\
  \bibinfo {author} {\bibfnamefont {F.}~\bibnamefont {Yan}},\ }\href {\doibase
  10.1088/1751-8121/aa7638} {\bibfield  {journal} {\bibinfo  {journal} {J.
  Phys. A: Math. Theor.}\ }\textbf {\bibinfo {volume} {50}},\ \bibinfo {pages}
  {285301} (\bibinfo {year} {2017})}\BibitemShut {NoStop}%
\bibitem [{\citenamefont {Xi}\ and\ \citenamefont
  {Yuwen}(2019{\natexlab{a}})}]{xi_coherence_2019}%
  \BibitemOpen
  \bibfield  {author} {\bibinfo {author} {\bibfnamefont {Z.}~\bibnamefont
  {Xi}}\ and\ \bibinfo {author} {\bibfnamefont {S.}~\bibnamefont {Yuwen}},\
  }\href {\doibase 10.1103/PhysRevA.99.022340} {\bibfield  {journal} {\bibinfo
  {journal} {Phys. Rev. A}\ }\textbf {\bibinfo {volume} {99}},\ \bibinfo
  {pages} {022340} (\bibinfo {year} {2019}{\natexlab{a}})}\BibitemShut
  {NoStop}%
\bibitem [{\citenamefont {Xi}\ and\ \citenamefont
  {Yuwen}(2019{\natexlab{b}})}]{xi_epsilon-smooth_2019}%
  \BibitemOpen
  \bibfield  {author} {\bibinfo {author} {\bibfnamefont {Z.}~\bibnamefont
  {Xi}}\ and\ \bibinfo {author} {\bibfnamefont {S.}~\bibnamefont {Yuwen}},\
  }\href {\doibase 10.1103/PhysRevA.99.012308} {\bibfield  {journal} {\bibinfo
  {journal} {Phys. Rev. A}\ }\textbf {\bibinfo {volume} {99}},\ \bibinfo
  {pages} {012308} (\bibinfo {year} {2019}{\natexlab{b}})}\BibitemShut
  {NoStop}%
\bibitem [{\citenamefont {Zhao}\ \emph {et~al.}(2018)\citenamefont {Zhao},
  \citenamefont {Liu}, \citenamefont {Yuan}, \citenamefont {Chitambar},\ and\
  \citenamefont {Ma}}]{zhao_one-shot_2018}%
  \BibitemOpen
  \bibfield  {author} {\bibinfo {author} {\bibfnamefont {Q.}~\bibnamefont
  {Zhao}}, \bibinfo {author} {\bibfnamefont {Y.}~\bibnamefont {Liu}}, \bibinfo
  {author} {\bibfnamefont {X.}~\bibnamefont {Yuan}}, \bibinfo {author}
  {\bibfnamefont {E.}~\bibnamefont {Chitambar}}, \ and\ \bibinfo {author}
  {\bibfnamefont {X.}~\bibnamefont {Ma}},\ }\href {\doibase
  10.1103/PhysRevLett.120.070403} {\bibfield  {journal} {\bibinfo  {journal}
  {Phys. Rev. Lett.}\ }\textbf {\bibinfo {volume} {120}},\ \bibinfo {pages}
  {070403} (\bibinfo {year} {2018})}\BibitemShut {NoStop}%
\bibitem [{\citenamefont {Zhou}\ \emph {et~al.}(2017)\citenamefont {Zhou},
  \citenamefont {Zhao}, \citenamefont {Yuan},\ and\ \citenamefont
  {Ma}}]{zhou_polynomial_2017}%
  \BibitemOpen
  \bibfield  {author} {\bibinfo {author} {\bibfnamefont {Y.}~\bibnamefont
  {Zhou}}, \bibinfo {author} {\bibfnamefont {Q.}~\bibnamefont {Zhao}}, \bibinfo
  {author} {\bibfnamefont {X.}~\bibnamefont {Yuan}}, \ and\ \bibinfo {author}
  {\bibfnamefont {X.}~\bibnamefont {Ma}},\ }\href {\doibase
  10.1088/1367-2630/aa91fa} {\bibfield  {journal} {\bibinfo  {journal} {New J.
  Phys.}\ }\textbf {\bibinfo {volume} {19}},\ \bibinfo {pages} {123033}
  (\bibinfo {year} {2017})}\BibitemShut {NoStop}%
\bibitem [{\citenamefont {Streltsov}\ \emph {et~al.}(2016)\citenamefont
  {Streltsov}, \citenamefont {Chitambar}, \citenamefont {Rana}, \citenamefont
  {Bera}, \citenamefont {Winter},\ and\ \citenamefont
  {Lewenstein}}]{streltsov_entanglement_2016}%
  \BibitemOpen
  \bibfield  {author} {\bibinfo {author} {\bibfnamefont {A.}~\bibnamefont
  {Streltsov}}, \bibinfo {author} {\bibfnamefont {E.}~\bibnamefont
  {Chitambar}}, \bibinfo {author} {\bibfnamefont {S.}~\bibnamefont {Rana}},
  \bibinfo {author} {\bibfnamefont {M.~N.}\ \bibnamefont {Bera}}, \bibinfo
  {author} {\bibfnamefont {A.}~\bibnamefont {Winter}}, \ and\ \bibinfo {author}
  {\bibfnamefont {M.}~\bibnamefont {Lewenstein}},\ }\href {\doibase
  10.1103/PhysRevLett.116.240405} {\bibfield  {journal} {\bibinfo  {journal}
  {Phys. Rev. Lett.}\ }\textbf {\bibinfo {volume} {116}},\ \bibinfo {pages}
  {240405} (\bibinfo {year} {2016})}\BibitemShut {NoStop}%
\bibitem [{\citenamefont {Chitambar}\ \emph {et~al.}(2016)\citenamefont
  {Chitambar}, \citenamefont {Streltsov}, \citenamefont {Rana}, \citenamefont
  {Bera}, \citenamefont {Adesso},\ and\ \citenamefont
  {Lewenstein}}]{chitambar_assisted_2016}%
  \BibitemOpen
  \bibfield  {author} {\bibinfo {author} {\bibfnamefont {E.}~\bibnamefont
  {Chitambar}}, \bibinfo {author} {\bibfnamefont {A.}~\bibnamefont
  {Streltsov}}, \bibinfo {author} {\bibfnamefont {S.}~\bibnamefont {Rana}},
  \bibinfo {author} {\bibfnamefont {M.~N.}\ \bibnamefont {Bera}}, \bibinfo
  {author} {\bibfnamefont {G.}~\bibnamefont {Adesso}}, \ and\ \bibinfo {author}
  {\bibfnamefont {M.}~\bibnamefont {Lewenstein}},\ }\href {\doibase
  10.1103/PhysRevLett.116.070402} {\bibfield  {journal} {\bibinfo  {journal}
  {Phys. Rev. Lett.}\ }\textbf {\bibinfo {volume} {116}},\ \bibinfo {pages}
  {070402} (\bibinfo {year} {2016})}\BibitemShut {NoStop}%
\bibitem [{\citenamefont {Streltsov}\ \emph {et~al.}(2017)\citenamefont
  {Streltsov}, \citenamefont {Rana}, \citenamefont {Bera},\ and\ \citenamefont
  {Lewenstein}}]{streltsov_towards_2017}%
  \BibitemOpen
  \bibfield  {author} {\bibinfo {author} {\bibfnamefont {A.}~\bibnamefont
  {Streltsov}}, \bibinfo {author} {\bibfnamefont {S.}~\bibnamefont {Rana}},
  \bibinfo {author} {\bibfnamefont {M.~N.}\ \bibnamefont {Bera}}, \ and\
  \bibinfo {author} {\bibfnamefont {M.}~\bibnamefont {Lewenstein}},\ }\href
  {\doibase 10.1103/PhysRevX.7.011024} {\bibfield  {journal} {\bibinfo
  {journal} {Phys. Rev. X}\ }\textbf {\bibinfo {volume} {7}},\ \bibinfo {pages}
  {011024} (\bibinfo {year} {2017})}\BibitemShut {NoStop}%
\bibitem [{\citenamefont {Chitambar}\ and\ \citenamefont
  {Hsieh}(2016)}]{Chitambar_coherence-entanglement2016}%
  \BibitemOpen
  \bibfield  {author} {\bibinfo {author} {\bibfnamefont {E.}~\bibnamefont
  {Chitambar}}\ and\ \bibinfo {author} {\bibfnamefont {M.-H.}\ \bibnamefont
  {Hsieh}},\ }\href {\doibase 10.1103/PhysRevLett.117.020402} {\bibfield
  {journal} {\bibinfo  {journal} {Phys. Rev. Lett.}\ }\textbf {\bibinfo
  {volume} {117}},\ \bibinfo {pages} {020402} (\bibinfo {year}
  {2016})}\BibitemShut {NoStop}%
\bibitem [{\citenamefont {Girolami}\ and\ \citenamefont
  {Yadin}(2017)}]{Girolami2017}%
  \BibitemOpen
  \bibfield  {author} {\bibinfo {author} {\bibfnamefont {D.}~\bibnamefont
  {Girolami}}\ and\ \bibinfo {author} {\bibfnamefont {B.}~\bibnamefont
  {Yadin}},\ }\href {\doibase 10.3390/e19030124} {\bibfield  {journal}
  {\bibinfo  {journal} {Entropy}\ }\textbf {\bibinfo {volume} {19}} (\bibinfo
  {year} {2017}),\ 10.3390/e19030124}\BibitemShut {NoStop}%
\bibitem [{\citenamefont {Ma}\ \emph {et~al.}(2016)\citenamefont {Ma},
  \citenamefont {Yadin}, \citenamefont {Girolami}, \citenamefont {Vedral},\
  and\ \citenamefont {Gu}}]{ma_converting_2016}%
  \BibitemOpen
  \bibfield  {author} {\bibinfo {author} {\bibfnamefont {J.}~\bibnamefont
  {Ma}}, \bibinfo {author} {\bibfnamefont {B.}~\bibnamefont {Yadin}}, \bibinfo
  {author} {\bibfnamefont {D.}~\bibnamefont {Girolami}}, \bibinfo {author}
  {\bibfnamefont {V.}~\bibnamefont {Vedral}}, \ and\ \bibinfo {author}
  {\bibfnamefont {M.}~\bibnamefont {Gu}},\ }\href {\doibase
  10.1103/PhysRevLett.116.160407} {\bibfield  {journal} {\bibinfo  {journal}
  {Phys. Rev. Lett.}\ }\textbf {\bibinfo {volume} {116}},\ \bibinfo {pages}
  {160407} (\bibinfo {year} {2016})}\BibitemShut {NoStop}%
\bibitem [{\citenamefont {Killoran}\ \emph {et~al.}(2016)\citenamefont
  {Killoran}, \citenamefont {Steinhoff},\ and\ \citenamefont
  {Plenio}}]{killoran_converting_2016}%
  \BibitemOpen
  \bibfield  {author} {\bibinfo {author} {\bibfnamefont {N.}~\bibnamefont
  {Killoran}}, \bibinfo {author} {\bibfnamefont {F.~E.~S.}\ \bibnamefont
  {Steinhoff}}, \ and\ \bibinfo {author} {\bibfnamefont {M.~B.}\ \bibnamefont
  {Plenio}},\ }\href {\doibase 10.1103/PhysRevLett.116.080402} {\bibfield
  {journal} {\bibinfo  {journal} {Phys. Rev. Lett.}\ }\textbf {\bibinfo
  {volume} {116}},\ \bibinfo {pages} {080402} (\bibinfo {year}
  {2016})}\BibitemShut {NoStop}%
\bibitem [{\citenamefont {Wang}\ \emph {et~al.}(2017)\citenamefont {Wang},
  \citenamefont {Tang}, \citenamefont {Wei}, \citenamefont {Yu}, \citenamefont
  {Ke}, \citenamefont {Xu}, \citenamefont {Li},\ and\ \citenamefont
  {Guo}}]{wang_directly_2017}%
  \BibitemOpen
  \bibfield  {author} {\bibinfo {author} {\bibfnamefont {Y.-T.}\ \bibnamefont
  {Wang}}, \bibinfo {author} {\bibfnamefont {J.-S.}\ \bibnamefont {Tang}},
  \bibinfo {author} {\bibfnamefont {Z.-Y.}\ \bibnamefont {Wei}}, \bibinfo
  {author} {\bibfnamefont {S.}~\bibnamefont {Yu}}, \bibinfo {author}
  {\bibfnamefont {Z.-J.}\ \bibnamefont {Ke}}, \bibinfo {author} {\bibfnamefont
  {X.-Y.}\ \bibnamefont {Xu}}, \bibinfo {author} {\bibfnamefont {C.-F.}\
  \bibnamefont {Li}}, \ and\ \bibinfo {author} {\bibfnamefont {G.-C.}\
  \bibnamefont {Guo}},\ }\href {\doibase 10.1103/PhysRevLett.118.020403}
  {\bibfield  {journal} {\bibinfo  {journal} {Phys. Rev. Lett.}\ }\textbf
  {\bibinfo {volume} {118}},\ \bibinfo {pages} {020403} (\bibinfo {year}
  {2017})}\BibitemShut {NoStop}%
\bibitem [{\citenamefont {Zheng}\ \emph {et~al.}(2018)\citenamefont {Zheng},
  \citenamefont {Ma}, \citenamefont {Wang}, \citenamefont {Fei},\ and\
  \citenamefont {Peng}}]{zheng_experimental_2018}%
  \BibitemOpen
  \bibfield  {author} {\bibinfo {author} {\bibfnamefont {W.}~\bibnamefont
  {Zheng}}, \bibinfo {author} {\bibfnamefont {Z.}~\bibnamefont {Ma}}, \bibinfo
  {author} {\bibfnamefont {H.}~\bibnamefont {Wang}}, \bibinfo {author}
  {\bibfnamefont {S.-M.}\ \bibnamefont {Fei}}, \ and\ \bibinfo {author}
  {\bibfnamefont {X.}~\bibnamefont {Peng}},\ }\href {\doibase
  10.1103/PhysRevLett.120.230504} {\bibfield  {journal} {\bibinfo  {journal}
  {Phys. Rev. Lett.}\ }\textbf {\bibinfo {volume} {120}},\ \bibinfo {pages}
  {230504} (\bibinfo {year} {2018})}\BibitemShut {NoStop}%
\bibitem [{\citenamefont {Smith}\ \emph {et~al.}(2017)\citenamefont {Smith},
  \citenamefont {Smolin}, \citenamefont {Yuan}, \citenamefont {Zhao},
  \citenamefont {Girolami},\ and\ \citenamefont {Ma}}]{smith_quantifying_2017}%
  \BibitemOpen
  \bibfield  {author} {\bibinfo {author} {\bibfnamefont {G.}~\bibnamefont
  {Smith}}, \bibinfo {author} {\bibfnamefont {J.~A.}\ \bibnamefont {Smolin}},
  \bibinfo {author} {\bibfnamefont {X.}~\bibnamefont {Yuan}}, \bibinfo {author}
  {\bibfnamefont {Q.}~\bibnamefont {Zhao}}, \bibinfo {author} {\bibfnamefont
  {D.}~\bibnamefont {Girolami}}, \ and\ \bibinfo {author} {\bibfnamefont
  {X.}~\bibnamefont {Ma}},\ }\href@noop {} {\bibfield  {journal} {\bibinfo
  {journal} {ArXiv170709928 Quant-Ph}\ } (\bibinfo {year} {2017})},\ \Eprint
  {http://arxiv.org/abs/1707.09928} {arXiv:1707.09928 [quant-ph]} \BibitemShut
  {NoStop}%
\bibitem [{\citenamefont {Zhang}\ \emph {et~al.}(2018)\citenamefont {Zhang},
  \citenamefont {Liu}, \citenamefont {Yu},\ and\ \citenamefont
  {Tong}}]{zhang_estimating_2018}%
  \BibitemOpen
  \bibfield  {author} {\bibinfo {author} {\bibfnamefont {D.-J.}\ \bibnamefont
  {Zhang}}, \bibinfo {author} {\bibfnamefont {C.~L.}\ \bibnamefont {Liu}},
  \bibinfo {author} {\bibfnamefont {X.-D.}\ \bibnamefont {Yu}}, \ and\ \bibinfo
  {author} {\bibfnamefont {D.~M.}\ \bibnamefont {Tong}},\ }\href {\doibase
  10.1103/PhysRevLett.120.170501} {\bibfield  {journal} {\bibinfo  {journal}
  {Phys. Rev. Lett.}\ }\textbf {\bibinfo {volume} {120}},\ \bibinfo {pages}
  {170501} (\bibinfo {year} {2018})}\BibitemShut {NoStop}%
\bibitem [{\citenamefont {Yu}\ and\ \citenamefont
  {G{\"u}hne}(2019)}]{yu_detecting_2019}%
  \BibitemOpen
  \bibfield  {author} {\bibinfo {author} {\bibfnamefont {X.-D.}\ \bibnamefont
  {Yu}}\ and\ \bibinfo {author} {\bibfnamefont {O.}~\bibnamefont {G{\"u}hne}},\
  }\href {\doibase 10.1103/PhysRevA.99.062310} {\bibfield  {journal} {\bibinfo
  {journal} {Phys. Rev. A}\ }\textbf {\bibinfo {volume} {99}},\ \bibinfo
  {pages} {062310} (\bibinfo {year} {2019})}\BibitemShut {NoStop}%
\bibitem [{\citenamefont {Gottesman}(1996)}]{Gottesman1996}%
  \BibitemOpen
  \bibfield  {author} {\bibinfo {author} {\bibfnamefont {D.}~\bibnamefont
  {Gottesman}},\ }\href {\doibase 10.1103/PhysRevA.54.1862} {\bibfield
  {journal} {\bibinfo  {journal} {Phys. Rev. A}\ }\textbf {\bibinfo {volume}
  {54}},\ \bibinfo {pages} {1862} (\bibinfo {year} {1996})}\BibitemShut
  {NoStop}%
\bibitem [{\citenamefont {Gottesman}(1997)}]{GottesmanPhD1997}%
  \BibitemOpen
  \bibfield  {author} {\bibinfo {author} {\bibfnamefont {D.}~\bibnamefont
  {Gottesman}},\ }\emph {\bibinfo {title} {Stabilizer Codes and Quantum Error
  Correction}},\ \href@noop {} {Ph.D. thesis},\ \bibinfo  {school} {California
  Institute of Technology.} (\bibinfo {year} {1997})\BibitemShut {NoStop}%
\bibitem [{\citenamefont {Winter}\ and\ \citenamefont
  {Yang}(2016)}]{winter_operational_2016}%
  \BibitemOpen
  \bibfield  {author} {\bibinfo {author} {\bibfnamefont {A.}~\bibnamefont
  {Winter}}\ and\ \bibinfo {author} {\bibfnamefont {D.}~\bibnamefont {Yang}},\
  }\href {\doibase 10.1103/PhysRevLett.116.120404} {\bibfield  {journal}
  {\bibinfo  {journal} {Phys. Rev. Lett.}\ }\textbf {\bibinfo {volume} {116}},\
  \bibinfo {pages} {120404} (\bibinfo {year} {2016})}\BibitemShut {NoStop}%
\bibitem [{\citenamefont {Zhao}\ \emph {et~al.}(2019)\citenamefont {Zhao},
  \citenamefont {Liu}, \citenamefont {Yuan}, \citenamefont {Chitambar},\ and\
  \citenamefont {Winter}}]{zhao2019one}%
  \BibitemOpen
  \bibfield  {author} {\bibinfo {author} {\bibfnamefont {Q.}~\bibnamefont
  {Zhao}}, \bibinfo {author} {\bibfnamefont {Y.}~\bibnamefont {Liu}}, \bibinfo
  {author} {\bibfnamefont {X.}~\bibnamefont {Yuan}}, \bibinfo {author}
  {\bibfnamefont {E.}~\bibnamefont {Chitambar}}, \ and\ \bibinfo {author}
  {\bibfnamefont {A.}~\bibnamefont {Winter}},\ }\href@noop {} {\bibfield
  {journal} {\bibinfo  {journal} {IEEE Transactions on Information Theory}\
  }\textbf {\bibinfo {volume} {65}},\ \bibinfo {pages} {6441} (\bibinfo {year}
  {2019})}\BibitemShut {NoStop}%
\bibitem [{\citenamefont {Yuan}\ \emph {et~al.}(2015)\citenamefont {Yuan},
  \citenamefont {Zhou}, \citenamefont {Cao},\ and\ \citenamefont
  {Ma}}]{yuan_intrinsic_2015}%
  \BibitemOpen
  \bibfield  {author} {\bibinfo {author} {\bibfnamefont {X.}~\bibnamefont
  {Yuan}}, \bibinfo {author} {\bibfnamefont {H.}~\bibnamefont {Zhou}}, \bibinfo
  {author} {\bibfnamefont {Z.}~\bibnamefont {Cao}}, \ and\ \bibinfo {author}
  {\bibfnamefont {X.}~\bibnamefont {Ma}},\ }\href {\doibase
  10.1103/PhysRevA.92.022124} {\bibfield  {journal} {\bibinfo  {journal} {Phys.
  Rev. A}\ }\textbf {\bibinfo {volume} {92}},\ \bibinfo {pages} {022124}
  (\bibinfo {year} {2015})}\BibitemShut {NoStop}%
\bibitem [{\citenamefont {Hayashi}\ and\ \citenamefont
  {Zhu}(2018)}]{hayashi2018secure}%
  \BibitemOpen
  \bibfield  {author} {\bibinfo {author} {\bibfnamefont {M.}~\bibnamefont
  {Hayashi}}\ and\ \bibinfo {author} {\bibfnamefont {H.}~\bibnamefont {Zhu}},\
  }\href {\doibase 10.1103/PhysRevA.97.012302} {\bibfield  {journal} {\bibinfo
  {journal} {Phys. Rev. A}\ }\textbf {\bibinfo {volume} {97}},\ \bibinfo
  {pages} {012302} (\bibinfo {year} {2018})}\BibitemShut {NoStop}%
\bibitem [{\citenamefont {Yuan}\ \emph {et~al.}(2019)\citenamefont {Yuan},
  \citenamefont {Zhao}, \citenamefont {Girolami},\ and\ \citenamefont
  {Ma}}]{yuan2019quantum}%
  \BibitemOpen
  \bibfield  {author} {\bibinfo {author} {\bibfnamefont {X.}~\bibnamefont
  {Yuan}}, \bibinfo {author} {\bibfnamefont {Q.}~\bibnamefont {Zhao}}, \bibinfo
  {author} {\bibfnamefont {D.}~\bibnamefont {Girolami}}, \ and\ \bibinfo
  {author} {\bibfnamefont {X.}~\bibnamefont {Ma}},\ }\href@noop {} {\bibfield
  {journal} {\bibinfo  {journal} {Advanced Quantum Technologies}\ }\textbf
  {\bibinfo {volume} {2}},\ \bibinfo {pages} {1900053} (\bibinfo {year}
  {2019})}\BibitemShut {NoStop}%
\bibitem [{\citenamefont {Rodr{\'\i}guez-Rosario}\ \emph
  {et~al.}(2013)\citenamefont {Rodr{\'\i}guez-Rosario}, \citenamefont
  {Frauenheim},\ and\ \citenamefont
  {Aspuru-Guzik}}]{rodriguez2013thermodynamics}%
  \BibitemOpen
  \bibfield  {author} {\bibinfo {author} {\bibfnamefont {C.~A.}\ \bibnamefont
  {Rodr{\'\i}guez-Rosario}}, \bibinfo {author} {\bibfnamefont {T.}~\bibnamefont
  {Frauenheim}}, \ and\ \bibinfo {author} {\bibfnamefont {A.}~\bibnamefont
  {Aspuru-Guzik}},\ }\href@noop {} {\bibfield  {journal} {\bibinfo  {journal}
  {arXiv preprint arXiv:1308.1245}\ } (\bibinfo {year} {2013})}\BibitemShut
  {NoStop}%
\bibitem [{\citenamefont {Cicalese}\ and\ \citenamefont
  {Vaccaro}(2002)}]{cicalese_supermodularity_2002}%
  \BibitemOpen
  \bibfield  {author} {\bibinfo {author} {\bibfnamefont {F.}~\bibnamefont
  {Cicalese}}\ and\ \bibinfo {author} {\bibfnamefont {U.}~\bibnamefont
  {Vaccaro}},\ }\href {\doibase 10.1109/18.992785} {\bibfield  {journal}
  {\bibinfo  {journal} {IEEE Trans. Inf. Theory}\ }\textbf {\bibinfo {volume}
  {48}},\ \bibinfo {pages} {933} (\bibinfo {year} {2002})}\BibitemShut
  {NoStop}%
\bibitem [{\citenamefont {Nielsen}\ and\ \citenamefont
  {Chuang}(2010)}]{nielsen_quantum_2010}%
  \BibitemOpen
  \bibfield  {author} {\bibinfo {author} {\bibfnamefont {M.~A.}\ \bibnamefont
  {Nielsen}}\ and\ \bibinfo {author} {\bibfnamefont {I.~L.}\ \bibnamefont
  {Chuang}},\ }\href@noop {} {\emph {\bibinfo {title} {Quantum Computation and
  Quantum Information}}},\ \bibinfo {edition} {10th}\ ed.\ (\bibinfo
  {publisher} {{Cambridge University Press}},\ \bibinfo {address} {{Cambridge ;
  New York}},\ \bibinfo {year} {2010})\BibitemShut {NoStop}%
\bibitem [{\citenamefont {D\"ur}\ \emph {et~al.}(2003)\citenamefont {D\"ur},
  \citenamefont {Aschauer},\ and\ \citenamefont {Briegel}}]{Dur2003}%
  \BibitemOpen
  \bibfield  {author} {\bibinfo {author} {\bibfnamefont {W.}~\bibnamefont
  {D\"ur}}, \bibinfo {author} {\bibfnamefont {H.}~\bibnamefont {Aschauer}}, \
  and\ \bibinfo {author} {\bibfnamefont {H.-J.}\ \bibnamefont {Briegel}},\
  }\href {\doibase 10.1103/PhysRevLett.91.107903} {\bibfield  {journal}
  {\bibinfo  {journal} {Phys. Rev. Lett.}\ }\textbf {\bibinfo {volume} {91}},\
  \bibinfo {pages} {107903} (\bibinfo {year} {2003})}\BibitemShut {NoStop}%
\bibitem [{\citenamefont {G\"uhne}\ \emph {et~al.}(2011)\citenamefont
  {G\"uhne}, \citenamefont {Jungnitsch}, \citenamefont {Moroder},\ and\
  \citenamefont {Weinstein}}]{Otfried2011}%
  \BibitemOpen
  \bibfield  {author} {\bibinfo {author} {\bibfnamefont {O.}~\bibnamefont
  {G\"uhne}}, \bibinfo {author} {\bibfnamefont {B.}~\bibnamefont {Jungnitsch}},
  \bibinfo {author} {\bibfnamefont {T.}~\bibnamefont {Moroder}}, \ and\
  \bibinfo {author} {\bibfnamefont {Y.~S.}\ \bibnamefont {Weinstein}},\ }\href
  {\doibase 10.1103/PhysRevA.84.052319} {\bibfield  {journal} {\bibinfo
  {journal} {Phys. Rev. A}\ }\textbf {\bibinfo {volume} {84}},\ \bibinfo
  {pages} {052319} (\bibinfo {year} {2011})}\BibitemShut {NoStop}%
\bibitem [{\citenamefont {Schlingemann}(2001)}]{schlingemann_stabilizer_2001}%
  \BibitemOpen
  \bibfield  {author} {\bibinfo {author} {\bibfnamefont {D.}~\bibnamefont
  {Schlingemann}},\ }\href@noop {} {\bibfield  {journal} {\bibinfo  {journal}
  {ArXivquant-Ph0111080}\ } (\bibinfo {year} {2001})},\ \Eprint
  {http://arxiv.org/abs/quant-ph/0111080} {arXiv:quant-ph/0111080} \BibitemShut
  {NoStop}%
\bibitem [{\citenamefont {Yao}\ \emph {et~al.}(2015)\citenamefont {Yao},
  \citenamefont {Xiao}, \citenamefont {Ge},\ and\ \citenamefont
  {Sun}}]{yao_quantum_2015}%
  \BibitemOpen
  \bibfield  {author} {\bibinfo {author} {\bibfnamefont {Y.}~\bibnamefont
  {Yao}}, \bibinfo {author} {\bibfnamefont {X.}~\bibnamefont {Xiao}}, \bibinfo
  {author} {\bibfnamefont {L.}~\bibnamefont {Ge}}, \ and\ \bibinfo {author}
  {\bibfnamefont {C.~P.}\ \bibnamefont {Sun}},\ }\href {\doibase
  10.1103/PhysRevA.92.022112} {\bibfield  {journal} {\bibinfo  {journal} {Phys.
  Rev. A}\ }\textbf {\bibinfo {volume} {92}},\ \bibinfo {pages} {022112}
  (\bibinfo {year} {2015})}\BibitemShut {NoStop}%
\bibitem [{\citenamefont {{de Vicente}}\ and\ \citenamefont
  {Streltsov}(2017)}]{de_vicente_genuine_2017}%
  \BibitemOpen
  \bibfield  {author} {\bibinfo {author} {\bibfnamefont {J.~I.}\ \bibnamefont
  {{de Vicente}}}\ and\ \bibinfo {author} {\bibfnamefont {A.}~\bibnamefont
  {Streltsov}},\ }\href {\doibase 10.1088/1751-8121/50/4/045301} {\bibfield
  {journal} {\bibinfo  {journal} {J. Phys. Math. Theor.}\ }\textbf {\bibinfo
  {volume} {50}},\ \bibinfo {pages} {045301} (\bibinfo {year}
  {2017})}\BibitemShut {NoStop}%
\bibitem [{\citenamefont {Liu}\ \emph {et~al.}(2017{\natexlab{b}})\citenamefont
  {Liu}, \citenamefont {Guo},\ and\ \citenamefont {Tong}}]{liu_enhancing_2017}%
  \BibitemOpen
  \bibfield  {author} {\bibinfo {author} {\bibfnamefont {C.~L.}\ \bibnamefont
  {Liu}}, \bibinfo {author} {\bibfnamefont {Y.-Q.}\ \bibnamefont {Guo}}, \ and\
  \bibinfo {author} {\bibfnamefont {D.~M.}\ \bibnamefont {Tong}},\ }\href
  {\doibase 10.1103/PhysRevA.96.062325} {\bibfield  {journal} {\bibinfo
  {journal} {Phys. Rev. A}\ }\textbf {\bibinfo {volume} {96}},\ \bibinfo
  {pages} {062325} (\bibinfo {year} {2017}{\natexlab{b}})}\BibitemShut
  {NoStop}%
\bibitem [{\citenamefont {Liu}\ and\ \citenamefont
  {Zhou}(2019)}]{liu_deterministic_2019}%
  \BibitemOpen
  \bibfield  {author} {\bibinfo {author} {\bibfnamefont {C.~L.}\ \bibnamefont
  {Liu}}\ and\ \bibinfo {author} {\bibfnamefont {D.~L.}\ \bibnamefont {Zhou}},\
  }\href {\doibase 10.1103/PhysRevLett.123.070402} {\bibfield  {journal}
  {\bibinfo  {journal} {Phys. Rev. Lett.}\ }\textbf {\bibinfo {volume} {123}},\
  \bibinfo {pages} {070402} (\bibinfo {year} {2019})}\BibitemShut {NoStop}%
\bibitem [{\citenamefont {Kim}\ \emph {et~al.}(2006)\citenamefont {Kim},
  \citenamefont {Fiorentino},\ and\ \citenamefont
  {Wong}}]{Kim2006_entangledphoton}%
  \BibitemOpen
  \bibfield  {author} {\bibinfo {author} {\bibfnamefont {T.}~\bibnamefont
  {Kim}}, \bibinfo {author} {\bibfnamefont {M.}~\bibnamefont {Fiorentino}}, \
  and\ \bibinfo {author} {\bibfnamefont {F.~N.~C.}\ \bibnamefont {Wong}},\
  }\href {\doibase 10.1103/PhysRevA.73.012316} {\bibfield  {journal} {\bibinfo
  {journal} {Phys. Rev. A}\ }\textbf {\bibinfo {volume} {73}},\ \bibinfo
  {pages} {012316} (\bibinfo {year} {2006})}\BibitemShut {NoStop}%
\bibitem [{\citenamefont {Chen}\ \emph {et~al.}(2007)\citenamefont {Chen},
  \citenamefont {Li}, \citenamefont {Zhang}, \citenamefont {Chen},
  \citenamefont {Goebel}, \citenamefont {Chen}, \citenamefont {Mair},\ and\
  \citenamefont {Pan}}]{Kai2007}%
  \BibitemOpen
  \bibfield  {author} {\bibinfo {author} {\bibfnamefont {K.}~\bibnamefont
  {Chen}}, \bibinfo {author} {\bibfnamefont {C.-M.}\ \bibnamefont {Li}},
  \bibinfo {author} {\bibfnamefont {Q.}~\bibnamefont {Zhang}}, \bibinfo
  {author} {\bibfnamefont {Y.-A.}\ \bibnamefont {Chen}}, \bibinfo {author}
  {\bibfnamefont {A.}~\bibnamefont {Goebel}}, \bibinfo {author} {\bibfnamefont
  {S.}~\bibnamefont {Chen}}, \bibinfo {author} {\bibfnamefont {A.}~\bibnamefont
  {Mair}}, \ and\ \bibinfo {author} {\bibfnamefont {J.-W.}\ \bibnamefont
  {Pan}},\ }\href {\doibase 10.1103/PhysRevLett.99.120503} {\bibfield
  {journal} {\bibinfo  {journal} {Phys. Rev. Lett.}\ }\textbf {\bibinfo
  {volume} {99}},\ \bibinfo {pages} {120503} (\bibinfo {year}
  {2007})}\BibitemShut {NoStop}%
\bibitem [{\citenamefont {Vallone}\ \emph {et~al.}(2008)\citenamefont
  {Vallone}, \citenamefont {Pomarico}, \citenamefont {De~Martini},\ and\
  \citenamefont {Mataloni}}]{Vallone2008}%
  \BibitemOpen
  \bibfield  {author} {\bibinfo {author} {\bibfnamefont {G.}~\bibnamefont
  {Vallone}}, \bibinfo {author} {\bibfnamefont {E.}~\bibnamefont {Pomarico}},
  \bibinfo {author} {\bibfnamefont {F.}~\bibnamefont {De~Martini}}, \ and\
  \bibinfo {author} {\bibfnamefont {P.}~\bibnamefont {Mataloni}},\ }\href
  {\doibase 10.1103/PhysRevLett.100.160502} {\bibfield  {journal} {\bibinfo
  {journal} {Phys. Rev. Lett.}\ }\textbf {\bibinfo {volume} {100}},\ \bibinfo
  {pages} {160502} (\bibinfo {year} {2008})}\BibitemShut {NoStop}%
\bibitem [{\citenamefont {Gao}\ \emph {et~al.}(2010)\citenamefont {Gao},
  \citenamefont {Lu}, \citenamefont {Yao}, \citenamefont {Xu}, \citenamefont
  {G{\"u}hne}, \citenamefont {Goebel}, \citenamefont {Chen}, \citenamefont
  {Peng}, \citenamefont {Chen},\ and\ \citenamefont {Pan}}]{Weibo2010}%
  \BibitemOpen
  \bibfield  {author} {\bibinfo {author} {\bibfnamefont {W.-B.}\ \bibnamefont
  {Gao}}, \bibinfo {author} {\bibfnamefont {C.-Y.}\ \bibnamefont {Lu}},
  \bibinfo {author} {\bibfnamefont {X.-C.}\ \bibnamefont {Yao}}, \bibinfo
  {author} {\bibfnamefont {P.}~\bibnamefont {Xu}}, \bibinfo {author}
  {\bibfnamefont {O.}~\bibnamefont {G{\"u}hne}}, \bibinfo {author}
  {\bibfnamefont {A.}~\bibnamefont {Goebel}}, \bibinfo {author} {\bibfnamefont
  {Y.-A.}\ \bibnamefont {Chen}}, \bibinfo {author} {\bibfnamefont {C.-Z.}\
  \bibnamefont {Peng}}, \bibinfo {author} {\bibfnamefont {Z.-B.}\ \bibnamefont
  {Chen}}, \ and\ \bibinfo {author} {\bibfnamefont {J.-W.}\ \bibnamefont
  {Pan}},\ }\href {\doibase 10.1038/nphys1603} {\bibfield  {journal} {\bibinfo
  {journal} {Nature Physics}\ }\textbf {\bibinfo {volume} {6}},\ \bibinfo
  {pages} {331} (\bibinfo {year} {2010})}\BibitemShut {NoStop}%
\bibitem [{\citenamefont {T\'oth}\ and\ \citenamefont
  {G\"uhne}(2005)}]{Toth2005}%
  \BibitemOpen
  \bibfield  {author} {\bibinfo {author} {\bibfnamefont {G.}~\bibnamefont
  {T\'oth}}\ and\ \bibinfo {author} {\bibfnamefont {O.}~\bibnamefont
  {G\"uhne}},\ }\href {\doibase 10.1103/PhysRevA.72.022340} {\bibfield
  {journal} {\bibinfo  {journal} {Phys. Rev. A}\ }\textbf {\bibinfo {volume}
  {72}},\ \bibinfo {pages} {022340} (\bibinfo {year} {2005})}\BibitemShut
  {NoStop}%
\bibitem [{\citenamefont {Zhao}\ and\ \citenamefont
  {Zhou}(2020)}]{zhao2020constructing}%
  \BibitemOpen
  \bibfield  {author} {\bibinfo {author} {\bibfnamefont {Q.}~\bibnamefont
  {Zhao}}\ and\ \bibinfo {author} {\bibfnamefont {Y.}~\bibnamefont {Zhou}},\
  }\href@noop {} {\enquote {\bibinfo {title} {Constructing multipartite bell
  inequalities from stabilizers},}\ } (\bibinfo {year} {2020}),\ \Eprint
  {http://arxiv.org/abs/2002.01843} {arXiv:2002.01843 [quant-ph]} \BibitemShut
  {NoStop}%
\bibitem [{\citenamefont {Gottesman}(1999)}]{Gottesman1999}%
  \BibitemOpen
  \bibfield  {author} {\bibinfo {author} {\bibfnamefont {D.}~\bibnamefont
  {Gottesman}},\ }in\ \href@noop {} {\emph {\bibinfo {booktitle} {Quantum
  Computing and Quantum Communications}}},\ \bibinfo {editor} {edited by\
  \bibinfo {editor} {\bibfnamefont {C.~P.}\ \bibnamefont {Williams}}}\
  (\bibinfo  {publisher} {Springer Berlin Heidelberg},\ \bibinfo {address}
  {Berlin, Heidelberg},\ \bibinfo {year} {1999})\ pp.\ \bibinfo {pages}
  {302--313}\BibitemShut {NoStop}%
\bibitem [{\citenamefont {Hostens}\ \emph {et~al.}(2005)\citenamefont
  {Hostens}, \citenamefont {Dehaene},\ and\ \citenamefont
  {De~Moor}}]{Hostens2005}%
  \BibitemOpen
  \bibfield  {author} {\bibinfo {author} {\bibfnamefont {E.}~\bibnamefont
  {Hostens}}, \bibinfo {author} {\bibfnamefont {J.}~\bibnamefont {Dehaene}}, \
  and\ \bibinfo {author} {\bibfnamefont {B.}~\bibnamefont {De~Moor}},\ }\href
  {\doibase 10.1103/PhysRevA.71.042315} {\bibfield  {journal} {\bibinfo
  {journal} {Phys. Rev. A}\ }\textbf {\bibinfo {volume} {71}},\ \bibinfo
  {pages} {042315} (\bibinfo {year} {2005})}\BibitemShut {NoStop}%
\bibitem [{\citenamefont {Gheorghiu}(2014)}]{Gheorghiu2014}%
  \BibitemOpen
  \bibfield  {author} {\bibinfo {author} {\bibfnamefont {V.}~\bibnamefont
  {Gheorghiu}},\ }\href {\doibase
  https://doi.org/10.1016/j.physleta.2013.12.009} {\bibfield  {journal}
  {\bibinfo  {journal} {Physics Letters A}\ }\textbf {\bibinfo {volume}
  {378}},\ \bibinfo {pages} {505 } (\bibinfo {year} {2014})}\BibitemShut
  {NoStop}%
\bibitem [{\citenamefont {Pati}\ \emph {et~al.}(2015)\citenamefont {Pati},
  \citenamefont {Singh},\ and\ \citenamefont {Sinha}}]{Pati2015}%
  \BibitemOpen
  \bibfield  {author} {\bibinfo {author} {\bibfnamefont {A.~K.}\ \bibnamefont
  {Pati}}, \bibinfo {author} {\bibfnamefont {U.}~\bibnamefont {Singh}}, \ and\
  \bibinfo {author} {\bibfnamefont {U.}~\bibnamefont {Sinha}},\ }\href
  {\doibase 10.1103/PhysRevA.92.052120} {\bibfield  {journal} {\bibinfo
  {journal} {Phys. Rev. A}\ }\textbf {\bibinfo {volume} {92}},\ \bibinfo
  {pages} {052120} (\bibinfo {year} {2015})}\BibitemShut {NoStop}%
\bibitem [{\citenamefont {Nirala}\ \emph {et~al.}(2019)\citenamefont {Nirala},
  \citenamefont {Sahoo}, \citenamefont {Pati},\ and\ \citenamefont
  {Sinha}}]{Nirala2019}%
  \BibitemOpen
  \bibfield  {author} {\bibinfo {author} {\bibfnamefont {G.}~\bibnamefont
  {Nirala}}, \bibinfo {author} {\bibfnamefont {S.~N.}\ \bibnamefont {Sahoo}},
  \bibinfo {author} {\bibfnamefont {A.~K.}\ \bibnamefont {Pati}}, \ and\
  \bibinfo {author} {\bibfnamefont {U.}~\bibnamefont {Sinha}},\ }\href
  {\doibase 10.1103/PhysRevA.99.022111} {\bibfield  {journal} {\bibinfo
  {journal} {Phys. Rev. A}\ }\textbf {\bibinfo {volume} {99}},\ \bibinfo
  {pages} {022111} (\bibinfo {year} {2019})}\BibitemShut {NoStop}%
\bibitem [{\citenamefont {Dai}\ \emph {et~al.}(2020)\citenamefont {Dai},
  \citenamefont {Dong}, \citenamefont {Xu}, \citenamefont {You}, \citenamefont
  {Zhang},\ and\ \citenamefont {G\"uhne}}]{DaiPhysRevApplied2020}%
  \BibitemOpen
  \bibfield  {author} {\bibinfo {author} {\bibfnamefont {Y.}~\bibnamefont
  {Dai}}, \bibinfo {author} {\bibfnamefont {Y.}~\bibnamefont {Dong}}, \bibinfo
  {author} {\bibfnamefont {Z.}~\bibnamefont {Xu}}, \bibinfo {author}
  {\bibfnamefont {W.}~\bibnamefont {You}}, \bibinfo {author} {\bibfnamefont
  {C.}~\bibnamefont {Zhang}}, \ and\ \bibinfo {author} {\bibfnamefont
  {O.}~\bibnamefont {G\"uhne}},\ }\href {\doibase
  10.1103/PhysRevApplied.13.054022} {\bibfield  {journal} {\bibinfo  {journal}
  {Phys. Rev. Applied}\ }\textbf {\bibinfo {volume} {13}},\ \bibinfo {pages}
  {054022} (\bibinfo {year} {2020})}\BibitemShut {NoStop}%
\end{thebibliography}%

\end{document}